\documentclass[aps,twocolumn,groupedaddress,superscriptaddress,floatfix,pra]{revtex4-2}
\usepackage{amsmath}
\usepackage{mathtools}
\usepackage{amssymb}
\usepackage{graphicx}
\usepackage{textcomp}
\usepackage{bm}	
\usepackage{soul}

\usepackage[usenames,dvipsnames]{color}

\usepackage[braket, qm]{qcircuit}

\begin{document}

\title{Efficiency of optically pumping a quantum battery and a two-stroke heat engine}

\author{Tiago F. F. Santos}\email{Corresponding author: tffs@pos.if.ufrj.br}
\affiliation{Instituto de F\'isica, Universidade Federal do Rio de Janeiro, CP68528, Rio de Janeiro, Rio de Janeiro 21941-972, Brazil}

\author{Marcelo F. Santos}
\affiliation{Instituto de F\'isica, Universidade Federal do Rio de Janeiro, CP68528, Rio de Janeiro, Rio de Janeiro 21941-972, Brazil}

\pacs{xxxx, xxxx, xxxx}                                         
\date{\today}
\begin{abstract}
In this work, we study the efficiency of charging a quantum battery through optical pumping. The battery consists of a qutrit and it is connected to a natural thermal reservoir and an external coherent drive in the limit where its upper energy level can be adiabatically eliminated from the dynamics. In this scenario, the drive plus spontaneous emission optically pumps the intermediate energy level of the qutrit and the battery can be understood as being charged by an effective higher temperature reservoir that takes it out of equilibrium with the natural reservoir and stores useful energy in it. We also analyse the efficiency of using this battery and charging scheme as the work fluid of a two-stroke thermal machine. The thermal machine includes a fourth level through which work is extracted from the battery via a unitary transformation, therefore setting the limit of maximum efficiency of the machine.
\end{abstract}

\maketitle

\section{Introduction}
The growing interest in quantum technologies and the continuous miniaturization of current devices has made it essential to understand the thermodynamics laws on micro and nano-scales. It has also led to a growing research interest in devices where microscopic media are used as the work fluid of heat engines or as energy storing quantum batteries.

A quantum heat engine is a machine whose work fluid is a quantum system operating in or out of equilibrium. In that sense, any quantum system coherently and incoherently exchanging energy with its environment can be interpreted as such a machine. In fact, the first quantum dynamics viewed as a quantum heat engine was a three-level maser~\cite{ScovilDuBois}. Since then, many different analogues of classical heat engines operating in equivalents Carnot and Otto cycles have been studied~\cite{Quan2007, Quan2006, Quan2005, Bender2000, Arnaud2002, friedemann2005, Kosloff2006, Kosloff2003, Thomas2011, Peterson2019, Popescu2014, Kosloff2015, Klaers2017, Halpern2019, gio2019, shag2022, ptas2022}, as well as many other more general models~\cite{,ptas2022, Alicki1979, Kosloff1984, Singer2016, Kosloff1996, Kosloff1994, Varinder2020, Boukobza2019, Sourav2020, Menczel2020, Gosh2018, Levy2016, Dorfman2021, zou2017, V2020}. In most cases, the main goal is to define the efficiency and output power of the machines, their particular regimes of operations and if they present some quantum advantage when compared to their classical counterparts, when there is one. 

In parallel to that, another relevant energetic problem related to the development of new quantum technologies concerns the efficiency and input power of the charging of quantum batteries~\cite{QB}, namely quantum systems that can store energy to be later extracted as work. The charging of quantum batteries has been studied both for unitary dynamics~\cite{Alicki2012, Binder2015, Campaioli2017, Ferraro2018, Le2018, Andolina2019, yuyu2019, cres2020} as well as for incoherent processes~\cite{Farina2019, Tacchino2020, Barra2019, Primordian2019, M2020, Santos2019, Caravelli2021, Xu2021, Si2020, zak2021, FBarra2022}. In most cases, the target of investigation has been the enhancement of input power due to quantum properties such as coherence and correlation~\cite{karen2013, andolina2018, caravelli2020, sergi2020, liu2021, Mondal2022, Gyhm2022, Ghosh2022, Kang2022}, whereas in~\cite{Farina2019, Barra2019, mayo2022, Fbarra2020, Barra2022, ghosh2022, stefano2022} the authors also analyse the energetic efficiency of the charging process. Most of them, however, do not include in the calculation of the charging efficiency the eventual cost to produce anomalous energy sources such as structured of engineered reservoirs or ``classical'' drives. A notable exception is found in Ref.~\cite{stefano2022}, where the authors compute the thermodynamic cost of measurements used to help stabilizing the charging of the battery.

One very common example of engineered reservoir is found in the ubiquitous method of optically pumping energy into an atomic system. In general, optical pumping involves the combination of one or more external coherent driving fields and spontaneous emission to incoherently transfer atomic populations between lower energy levels with the mediation of a higher energy one. If the parameters are such that this higher energy level can be adiabatically eliminated from the dynamics, the overall time evolution can be approximately treated in the subspace of the lower energy levels and the optical pumping may eventually mimic a thermal reservoir. In particular, in the simplest scenario of the population transfer between a ground state to an excited level, the variation of internal energy of the system will be exactly equal to the heat pumped by this effective reservoir. This calculation, however, does not take into consideration the energetic cost to create such effective reservoir and may even lead to the false impression that the charging of the atomic system can be done with efficiency one. Furthermore, it does not reveal the thermodynamics limitations and costs to sustain such effective reservoir if the temperature of the environment surrounding the system increases.

Our goal in this work is to investigate these thermodynamic costs and limitations in this simplest case of optically pumping energy into an effective two-level quantum battery. We particularly focus on analyzing the efficiency and input power of the charging of this battery as we change parameters such as the energy separation of its levels, the pumping rate and the temperature of the outside environment. The battery consists of a qutrit that is coupled to an external work drive and to a standard thermal reservoir in a typical optical pumping setup (see Fig.~\ref{fig1}). We assume that the qutrit is initially in thermal equilibrium with the reservoir and it is driven out of equilibrium by the coherent external source, so that useful energy is stored in it. We analyze the efficiency and the stored power in this process taking into consideration the energy transferred by the external ``classical'' source as well as that exchanged with the thermal reservoir, in the limit where the upper energy level of the qutrit can be adiabatically eliminated from the dynamics~\cite{AE1,AE2}.

In the sequence, we use this battery as the quantum work fluid of the two-stroke thermal machine analyzed in~\cite{Tiago2021}. There, the battery is charged through a heat current established when reservoirs of distinct temperatures affect different transitions of a three-level atom and it is shown that the machine operates at the Otto cycle efficiency. However, this calculation does not consider the cost of preparing the effective reservoir. Here, we re-calculate the efficiency and output power of the machine considering that the effective reservoir is the result of an optical pumping process. The engine includes a fourth level and operates a cycle where energy is stored through the optical pumping and extracted by means of a unitary transformation.

The paper is organized as follows. In Sec.~\ref{sec:theory} we revisit the theory of quantum open system dynamics and thermodynamics and present our models of the quantum battery and the two-stroke quantum heat engine. We also show some algebraic results for particular regimes of operation of the machine. In Sec.~\ref{sec:results} we show and analyze the numerical results for both setups. Finally, in Sec.~\ref{sec:conclusion} we summarize and comment on the results and discuss possible future developments. Details of the calculations are reported in the Appendix

\section{Theoretical Background}\label{sec:theory}

\subsection{Open quantum system dynamics and thermodynamics}
Here we will assume that the interaction between the quantum system and the thermal reservoir is Markovian so that the dynamics is ruled by a master equation in the Lindblad form \cite{open} ($\hbar = k_B = 1$ in the following):
\begin{equation}\label{LME}
\dot{\rho} = -i[H(t),\rho] + \mathcal{L}(\rho).
\end{equation}
The first term on the right side accounts for the unitary part of the dynamics. The Hamiltonian reads $H(t) = H_0 + V(t)$, where $H_0$ is the free Hamiltonian of the system (e.g. non-degenerate three or four-level atoms) and $V(t)$ accounts for its coupling with external coherent sources (e.g. laser fields). The second term represents the non-unitary part of the dynamics, $\mathcal{L}(\rho)$,  given by
\begin{equation}\label{dissipative}
    \mathcal{L}(\rho) = \sum_j L_j(\rho) =\sum_j  \Gamma_j\left[J_j \rho J_j^{\dagger} - \frac{1}{2}\{J_j^{\dagger}J_j, \rho \}\right],
\end{equation}
where $\{A,B\} = AB + BA$, $\Gamma_j$ are the incoherent energy exchange rates and $J_j$ are the corresponding energy jump operators in the system. In this scenario and defining the internal energy of the system at time $t$ as the expectation value of its total energy, $U(t)=\textrm{Tr}\{ \rho(t) H(t)\}$, R. Alicki~\cite{Alicki1979} has defined the work, $W$, and the heat, $Q$, exchanged between the system and the reservoirs respectively by
\begin{equation}\label{alickiwork}
    W = \int_{t_0}^t dt' \operatorname{Tr}\{\rho(t') \dot{H}(t')\}
\end{equation}
and
\begin{align}\label{alickiheat}
    Q &= \int_{t_0}^t dt' \operatorname{Tr}\{\dot{\rho}(t') H(t')\} = \int_{t_0}^t dt' \operatorname{Tr}\{\mathcal{L}[\rho(t')] H(t')\}.
\end{align}
Note that these definitions automatically satisfy the first law of thermodynamics. Even though there has been a long discussion about the definitions of work and heat in quantum thermodynamics~\cite{felix2015, ken2018, B2020, wei2022}, for all the purposes of this paper, Alicki's are satisfactory.

In order to analyze the efficiency of the charging of our battery we need to compute its non-equilibrium Helmholtz free energy, $F(t)=U(t)-TS(t)$, where $T$ is the temperature of the reservoir in which the system is embedded and $S(t)$ is the von Neumann entropy of the system, $S(t) = - \operatorname{Tr}\{\rho(t) ln\left[\rho(t)\right]\}$. We also need to remember that in an isothermal and generally irreversible process, the maximum amount of work, $W_{ext}$, that can be extracted or stored in the battery is at most equal to a decrease or increase in its Helmholtz free energy, where the equality holds for reversible processes.

\subsection{Quantum battery model}\label{QBM}
\begin{figure}
  \centering
   \includegraphics[width=\columnwidth]{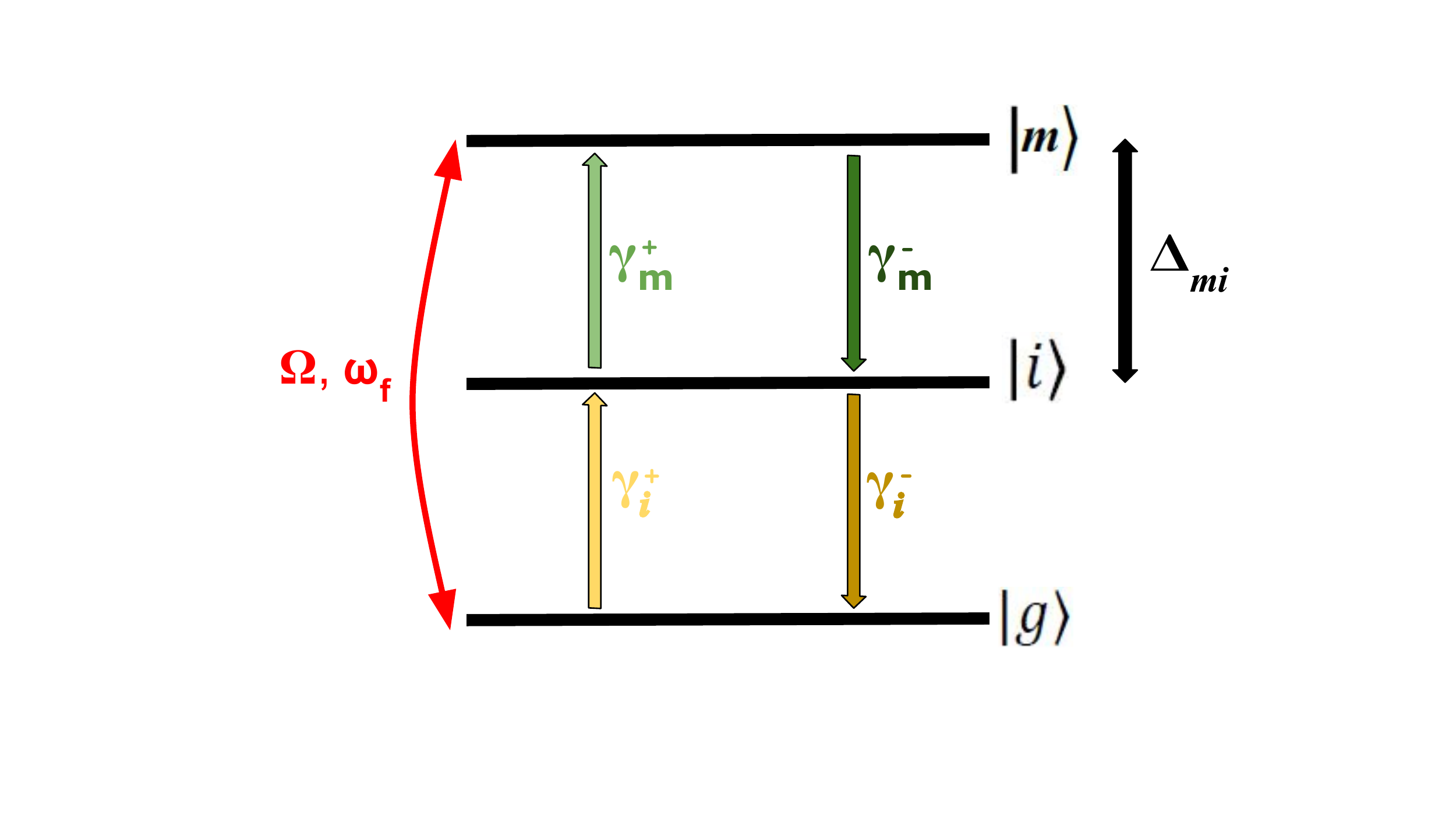}
   \caption{A three-level system in cascade configuration scheme. The transitions between the energy levels are induced by the coupling of the system to a thermal reservoir and to an external work drive that injects energy into the system. $\Delta_{mi} = E_m - E_i$.} \label{fig1}
\end{figure}

We consider a quantum battery (Fig. \ref{fig1}) that consists of a qutrit of levels $\ket{g}$, $\ket{i}$ and $\ket{m}$ ($E_g = 0 < E_i < E_m$) connected to an external work drive, $V_{in}(t)=\Omega \hskip 0.01cm \left(\sigma_{gm} e^{i \omega_f t} + \sigma_{mg} e^{-i \omega_f t}\right)$ ($\sigma_{kl}\equiv |k\rangle\langle l|$) that couples levels $\ket{g}$ and $\ket{m}$ and injects energy into the system, and to a thermal reservoir of finite temperature $T$ that generates a Markovian incoherent energy exchange with the system. The dynamics of the battery is given by Eqs.~\ref{LME} and~\ref{dissipative} where $\mathcal{L}(\rho)$ is decomposed in the following terms:
\begin{align}\label{lgm}
\mathcal{L}_{\gamma_{m}}[\rho(t)] &= \frac{\gamma_m^{+}}{2}\left(2 \sigma_{mi} \rho(t) \sigma_{im} - \{\sigma_{ii}, \rho(t)\}\right) \nonumber\\
&+\frac{\gamma_m^{-}}{2}\left(2 \sigma_{im} \rho(t) \sigma_{mi} - \{\sigma_{mm}, \rho(t)\}\right),
\end{align}
where $\gamma_m^{+} = \gamma_0^{m} \hskip 0.1cm \bar{n}_m$, $\gamma_m^{-} = \gamma_0^{m} \hskip 0.1cm (\bar{n}_m + 1)$ and $\bar{n}_m = \left(e^{\frac{E_m - E_i}{T}} - 1\right)^{-1}$, and
\begin{align}\label{lgi}
\mathcal{L}_{\gamma_{i}}[\rho(t)] &= \frac{\gamma_i^{+}}{2}\left(2 \sigma_{ig} \rho(t) \sigma_{gi} - \{\sigma_{gg}, \rho(t)\}\right) \nonumber\\
&+\frac{\gamma_i^{-}}{2}\left(2 \sigma_{gi} \rho(t) \sigma_{ig} - \{\sigma_{ii}, \rho(t)\}\right),
\end{align}
where $\gamma_i^{+} = \gamma_0^{i} \hskip 0.1cm \bar{n}_i$, $\gamma_i^{-} = \gamma_0^{i} \hskip 0.1cm (\bar{n}_i + 1)$ and $\bar{n}_i = \left(e^{\frac{E_i - E_g}{T}} - 1\right)^{-1}$. Here, $\gamma_0^m$ ($\gamma_0^i$) denotes the spontaneous emission rate for the transition $\ket{m} \rightarrow \ket{i}$ ($\ket{i} \rightarrow \ket{g}$). Note that we consider that the reservoir does not directly couples levels $|m\rangle$ and $|g\rangle$ or that, if it does, this coupling is very weak when compared to the others (two approximations commonly found in quantum optical systems).

We assume that the system is in contact with the natural reservoir at temperature T and, thus, it is initially prepared in a Gibbs state, $\rho(0) = e^{-\beta H_0}/Z$, where $H_0 = \omega_j \ket{E_j}\bra{E_j}$, $\beta = 1/T$ and $Z = \operatorname{Tr}(e^{-\beta H_0})$ is the partition function. Turning on $V_{in}(t)$ evolves the system into a non-equilibrium energy storing steady state, $\rho_{NESS}$, whose free energy is larger than that of $\rho(0)$ (see the Appendix for a detailed calculation of the dynamics). The efficiency of this process is given by $\eta_{pump} = \frac{\Delta F}{E_{in}}$, and its input power by $\mathcal{P}_{pump} = \frac{\Delta F}{\tau}$, where $E_{in}$ is the total energy injected into the system, $\tau$ is the time it takes for the system to reach $\rho_{NESS}$ from the initial Gibbs state $\rho(0)$, and $\Delta F = F_{NESS} - F(0)$ is the variation of Helmholtz free energy. There are two possible energy sources for the process: the external drive and the thermal reservoir. The first will always inject energy in the battery whereas the latter may or may not do so, depending on the parameters. In order to compute the total amount of injected energy, $E_{in}$, we need to consider
\begin{equation}
    E_{in} = Max\{W_{in}, W_{in} + Q_{\gamma_m}, W_{in} + Q_{\gamma_i}, W_{in} + Q\},
\end{equation}
where $W_{in}$ is the work done by the external source and $Q_{k}$ the heat injected by reservoir $k$. Taking $V_{in}(t)$ into consideration, the injected work is given by
\begin{equation}\label{w3n}
    W_{in} = i \Omega \omega_f \int_0^{\tau} dt \hskip 0.1cm [\varrho_{mg}(t) - \varrho_{gm}(t)],
\end{equation}
where $\varrho_{gm} = \langle g|e^{i H_0^{'} t} \rho e^{-i H_0^{'} t}|m\rangle$, $H_0^{'} = H_0 + \Delta \omega \ket{m}\bra{m}$ and $\Delta \omega = \omega_f - \omega_m$, whereas the heat components of the energy exchange are given by 
\begin{align}\label{qm3n}
   Q_{\gamma_m} &= -\Omega \frac{\gamma_m^{-}}{2} \int_0^{\tau} dt \hskip 0.1cm [\varrho_{mg}(t) + \varrho_{gm}(t)] \nonumber \\
   &+ (\omega_{m} - \omega_{i}) \int_0^{\tau} dt \hskip 0.1cm [\gamma_m^{+} \varrho_{ii}(t) - \gamma_m^{-} \varrho_{mm}(t)],
\end{align}
between levels $|m\rangle$ and $|i\rangle$, and 
\begin{align}\label{qi3n}
   Q_{\gamma_i} &= -\Omega \frac{\gamma_i^{+}}{2} \int_0^{\tau} dt \hskip 0.1cm [\varrho_{mg}(t) + \varrho_{gm}(t)] \nonumber \\
   &+ \omega_i \int_0^{\tau} dt \hskip 0.1cm [\gamma_i^{+} \varrho_{gg}(t) - \gamma_i^{-} \varrho_{ii}(t)],
\end{align}
between levels $|i\rangle$ and $|g\rangle$ (See Appendix for details). $Q =  Q_{\gamma_m} + Q_{\gamma_i}$ is the total heat exchange.

In this work, we are interested in the scenario where we can adiabatically eliminate level $\ket{m}$ from the dynamics. This limit is achieved if we consider that the decay rate from level $\ket{m}$ to $\ket{i}$ is much larger than the other rates involved in the dynamics, $\gamma_m^{-} \gg \gamma_m^{+}, \gamma_i^{\pm}, \Omega$. Note that $\gamma_m^{-} \gg \gamma_m^{+}$ means that $k_B T \gg E_m-E_i$, establishing either a maximum temperature or a minimum energy separation between levels $|m\rangle$ and $|i\rangle$ for the adiabatic elimination to work. Under these conditions, the population of level $|m\rangle$ becomes approximately stationary and very small when compared to the others, in a time scale that is much smaller than the effective changes experienced by the battery. In this case, the combination of the external work drive with spontaneous emission optically pumps level $|i\rangle$, i.e. population is effectively and incoherently transferred straight from level $|g\rangle$ to level $|i\rangle$, and Eq. \eqref{w3n} reads (See Appendix for details)
\begin{equation}\label{w3nea}
    W_{in} = p \omega_f \gamma_m^{-} \int_0^{\tau} dt \hskip 0.1cm\varrho_{gg}(t),
\end{equation}
where
\begin{equation}\label{p}
    p = \frac{4 \Omega^2}{\gamma_m^{-^{2}} + 4 \Delta \omega^2}
\end{equation}
is the pumping rate. For fixed values of $\Omega$ and $\gamma_m^{-}$, $p$ is maximized when $\Delta \omega = \omega_f-\omega_m = 0$, i.e., when the laser drives the $|g\rangle \rightarrow |m\rangle$ transition resonantly. As for the exchanged heat, Eqs.~\eqref{qm3n} and \eqref{qi3n} read (See Appendix for details)
\begin{align}
    Q_{\gamma_m} &\approx p \gamma_m^{-}(\omega_i - \omega_f) \int_0^{\tau} dt \hskip 0.1cm\varrho_{gg}(t)
\end{align}
and
\begin{align}
    Q_{\gamma_i} &= - p \gamma_i^{+} \Delta \omega \int_0^{\tau} dt \hskip 0.1cm\varrho_{gg}(t) \nonumber \\
    &+ \omega_i \int_0^{\tau} dt \hskip 0.1cm (\gamma_i^{+} \varrho_{gg}(t) - \gamma_i^{-} \varrho_{ii}(t)).
\end{align}
These are the equations ultimately used to compute $\eta_{pump}$ and $\mathcal{P}_{pump}$. Before advancing to the next section it is worth noting that the optical pumping plus the natural reservoir can be seen as an effective higher temperature reservoir acting on the $\{g\rangle,|i\rangle\}$ transition, given that the adiabatic elimination of level $|m\rangle$ produces an effective dynamics in this subspace, that corresponds to an extra term of the type $\mathcal{L}_{p}(\rho) = p\gamma^-_m(2\sigma_{ig} \rho \sigma_{gi} - \{\sigma_{gg},\rho\})$. This term adds to the ones produced by the natural reservoir, described by Eq.~\ref{lgi}, and unbalances the natural ratio $\frac{\gamma^-_i}{\gamma^+_i}=e^{\frac{E_i-E_g}{k_BT}}$ in order to produce an effective higher temperature $T_H=\frac{\hbar\omega_{i}}{k_B\textrm{ln}(\frac{\gamma^-_i}{p\gamma^-_m+\gamma_i^+})}$ in the subspace~\cite{ARRCMFS, MFSARRC, MFSDG}. This effective temperature can even be negative, if $p\gamma_m^- > \gamma_0^i$. The main goal of this paper is to analyse the energy cost of creating this effective reservoir.

\subsection{Quantum heat engine model} \label{QHE}
The proposed quantum heat engine is based on the version developed in~\cite{Tacchino2020} of a two-stage machine. The first stage uses an effective temperature difference to recharge the work fluid and the second stage, a unitary operation to extract work from it. Here, we will consider the battery described in the last section as the work fluid of the machine. 

\begin{figure}
  \centering
   \includegraphics[width=\columnwidth]{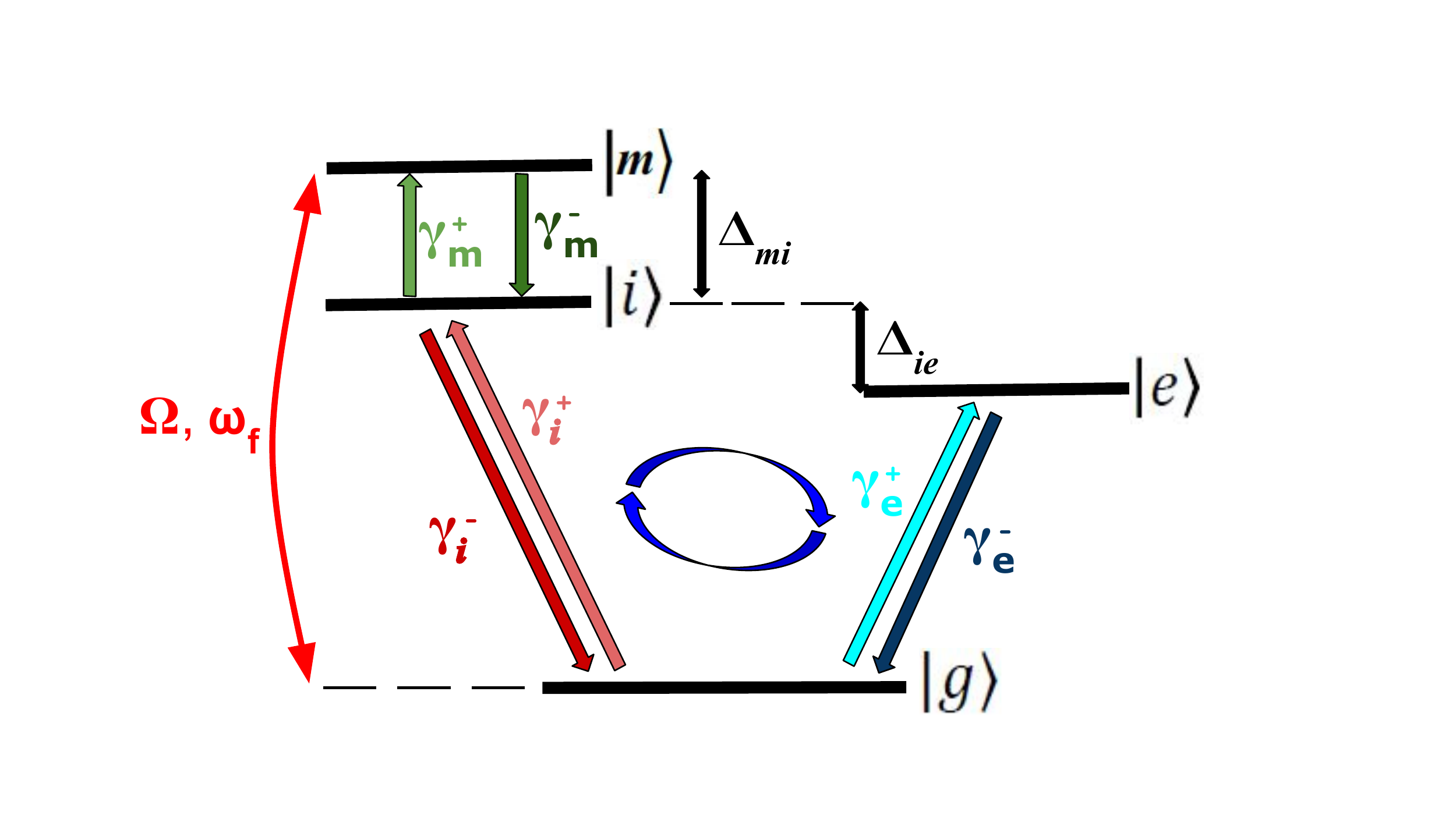}
   \caption{A four-level system as the working fluid. The transitions between the energy levels are induced by the coupling of the system to a thermal reservoir and to an external work drive that injects energy into the system. $\Delta_{mi} = E_m - E_i$ and $\Delta_{ie} = E_i - E_e$.} \label{fig2}
\end{figure}

The operation of the machine requires introducing a fourth level $|e\rangle$ into the system, whose energy $E_e$ lies in between those of levels $|g\rangle$ and $|i\rangle$ (see Fig. \ref{fig2}) - $E_g = 0 < E_e < E_i < E_m$. The $|g\rangle \rightarrow |e\rangle$ transition is in contact only with the natural reservoir at temperature $T$. When level $|m\rangle$ is adiabatically eliminated from the dynamics, the optical pumping of level $|i\rangle$, produced by $V_{in}(t)$, eventually inverts the population in the subspace $\{|e\rangle,|i\rangle\}$. This is equivalent to say that the machine operates under a temperature difference $T_H-T$ where $T_H$ ($T$) affects the $\{|g\rangle,\|i\rangle\}$ ($\{|g\rangle,\|e\rangle\}$) subspace. This difference in temperature establishes a heat flux through the system and if heat flows through the system for a long enough time it eventually produces states that are diagonal in the $H_0$ eigenbasis and given by $\rho_{OSS}= \sum_k r_k\sigma_{kk}$ ($k=g,e,i$), with $r_g > r_i >r_e$. Because the machine operates in closed cycles, we refer to these states as ``Operational Steady States'', hence the label OSS. Such states have positive ergotropy~\cite{Ergotropy}, $\mathcal{E}=(E_i-E_e)(r_i-r_e)$, meaning they store energy that can be extracted in the form of work by unitary operations. The largest population difference (largest value of ergotropy) is achieved asymptotically and corresponds to the most amount of energy that can be stored by the recharging stage. This particular OSS is called the ``Non-equilibrium steady state (NESS)'', from now on labelled $\rho_{NESS}$. Previous works~\cite{Tacchino2020, Tiago2021} have shown, however, that NESS is not necessarily the best regime for the operation of the machine. In general, intermediate charging times will be used, generating $\rho_{OSS}$ that carry less ergotropy but are much faster to recharge. The faster recharging time, from now on referred to as $\tau_r$, increases both the output power that depends on the inverse of the total duration of the cycle as well as the efficiency since less heat is lost to the reservoirs. To each $\tau_r$ there corresponds a particular state $\rho_{OSS}$.

The second stage of the machine, called the discharging stage, corresponds to a swap of the $|e\rangle$ and $|i\rangle$ populations, for the class of OSS here analyzed. Physically, such swap can be implemented by turning on a second external source, $V_{ext}(t)= \epsilon \hskip 0.1cm \left( \sigma_{ei} e^{i (\omega_i- \omega_e) t} + \sigma_{ie} e^{-i (\omega_i - \omega_e) t}\right)$ for a finite time $\tau_d = \frac{\pi}{2\epsilon}$, in what is known as a $\pi/2$ pulse. This pulse takes $\rho_{OSS}$ into a corresponding passive state~\cite{passive} of the form $\rho_{OSS}(\tau_d) = r_g|g\rangle\langle g|+r_i|e\rangle\langle e|+r_e|i\rangle \langle i|$, from which it can be recharged in the next cycle. Note that, by construction, the energy gained during the recharging stage balances out the energy extracted during the discharging stage so that $\Delta U_{cycle} = 0$ in a cycle, as expected.

From the dynamical point of view, taking level $|e\rangle$ into consideration adds extra non-unitary terms to the time evolution of the machine. Once again we will consider that the natural reservoir directly couples only levels $|e\rangle$ and $|g\rangle$ but does not act (or acts very weakly) in the $|e\rangle \rightarrow |i\rangle$ transition. This approximation simplifies the calculations with no loss of generality. The new terms read
\begin{align}\label{lge}
\mathcal{L}_{\gamma_{e}}[\rho(t)] &= \frac{\gamma_e^{+}}{2}\left(2 \sigma_{eg}\rho(t) \sigma_{ge} - \{\sigma_{gg}, \rho(t)\}\right) \nonumber\\
&+\frac{\gamma_e^{-}}{2}\left(2 \sigma_{ge} \rho(t) \sigma_{eg} - \{\sigma_{ee}, \rho(t)\}\right),
\end{align}
where $\gamma_e^{+} = \gamma_0^{e} \hskip 0.1cm \bar{n}_e$, $\gamma_e^{-} = \gamma_0^{e} \hskip 0.1cm (\bar{n}_e + 1)$, $\bar{n}_e = \left(e^{\frac{E_e - E_g}{T}} - 1\right)^{-1}$ and $\gamma_0^{e}$ is the spontaneous emission rate for the transition $\ket{e} \rightarrow \ket{g}$. Adiabatic elimination of level $|m\rangle$ also imposes $\gamma_m^{-} \gg  \gamma_{e}^{\pm}$.

At the end of one cycle, the work done on the system, $W_{in}$, the work performed by the machine, $W_{ext}$, and the heat exchanged between the system and the thermal reservoir, $Q$, are, once again, computed using Eqs.~\ref{alickiwork} and~\ref{alickiheat}. 

The efficiency of the machine is given by $\eta = -\frac{W_{ext}}{E_{in}}$,
where $E_{in}$ is the total energy injected in the system in one cycle, whereas its power output is $\mathcal{P} = -W_{ext}/\tau$.

In general, the energy exchanged by the machine as well as its efficiency and power output are obtained by numerically integrating in time different functions of the state of the system, similar to the ones obtained for the battery in Eqs.~\ref{w3n}-\ref{qi3n}. However, the calculations of these quantities are significantly simpler in the ideal operational regime of the machine, derived in previous works~\cite{Tacchino2020,Tiago2021}, that is achieved for $\tau_d \ll \tau_r$ and $\sum_j \gamma_j \tau \ll 1$. The first condition allows us to assume that the discharging stage is isentropic and isochoric, meaning that the extracted work is maximal, i.e. equal to the ergotropy stored in the system. The second condition establishes the short cycle operation, described in details in~\cite{Tacchino2020,Tiago2021} and characterized by a linear variation with time of the thermodynamics quantities and the minimization of the heat exchanged with the reservoirs. In this regime, the work injected in the machine is given by
\begin{align}\label{w16}
    W_{in}^{SC} &= \operatorname{Tr}[\tilde{\rho}\dot{V}_{in}(\tau)]\tau \nonumber \\
    &=\omega_f \hskip 0.08cm p \gamma_m^{-} \frac{\gamma_i^{-} + \gamma_e^{-}}{\kappa}\tau,
\end{align}
where $\kappa = 2\left(\Gamma_i^{+} + \gamma_e^{+}\right) + \gamma_i^{-} +  \gamma_e^{-}$, $\Gamma_i^{+} = \gamma_i^{+} + p \gamma_m^{-}$ and $p$ is given by Eq. \eqref{p}. The heat, on its turn, can be decomposed in the following three components:

\begin{align}
    Q_{\gamma_e}^{SC} &= \operatorname{Tr}\{\mathcal{L}_{\gamma_e}[\tilde{\rho}][H_0 + V_{in}(\tau)]\} \tau \nonumber \\
    &=\omega_e \frac{\gamma_e^{+} \gamma_i^{-} - \Gamma_i^{+} \gamma_e^{-}}{\kappa} \tau - p \gamma_e^{+} \Delta \omega \frac{\gamma_i^{-} + \gamma_e^{-}}{\kappa}\tau,
\end{align}
\begin{align}
    Q_{\gamma_i}^{SC} &= \operatorname{Tr}\{\mathcal{L}_{\gamma_i}[\tilde{\rho}][H_0 + V_{in}(\tau)]\} \tau \nonumber \\
    &=\omega_i \frac{\gamma_i^{+} \gamma_e^{-} - \gamma_i^{-}( \gamma_e^{+} + p \gamma_m^{-})}{\kappa} \tau \nonumber \\
    &- p \gamma_i^{+} \Delta \omega \frac{\gamma_i^{-} + \gamma_e^{-}}{\kappa}\tau,
\end{align}
and
\begin{align}\label{qm19}
    Q_{\gamma_m}^{SC} &=\operatorname{Tr}\{\mathcal{L}_{\gamma_m}[\tilde{\rho}][H_0 + V_{in}(\tau)]\}\tau \nonumber \\
    &= (\omega_i - \omega_f) \hskip 0.08cm p \gamma_m^{-} \frac{\gamma_i^{-} + \gamma_e^{-}}{\kappa}\tau.
\end{align}
To obtain the results in Eqs.(\ref{w16} - \ref{qm19}) we use that $\varrho = e^{i H_0^{'} t} \rho e^{-i H_0^{'} t}$, whereas $\tilde{\rho} = U \rho U^{-1}$ is the corresponding passive state of the system in this limit of operation, where $U$ is a unitary transformation that swaps the populations of levels $\ket{e}$ and $\ket{i}$. (See Appendix for details). Finally, the ergotropy stored in the system can also be algebraically computed and it is given by
\begin{equation}\label{EFFSC}
    \mathcal{E}_{SC} = (\omega_i - \omega_e) \frac{\Gamma_i^+ \gamma_e^{-} - \gamma_i^{-} \gamma_e^{+}}{\kappa} \tau.
\end{equation}
Note that all the quantities above are proportional to the cycle duration $\tau$, and for the ergotropy to be positive it is necessary that $\frac{\Gamma_i^{+}}{\gamma_i{-}} > \frac{\gamma_e^{+}}{\gamma_e{-}}$. Also note that if this condition is not fulfilled, a similar scheme can still be used as a refrigerator~\cite{R1, R2, R3, R4}. Finally note that the linear dependence on time of both ergotropy and injected energy makes both the efficiency of the short cycle, $\eta_{SC} = \mathcal{E}_{SC}/E_{in}^{SC}$ and its power output $\mathcal{P}_{SC} = \mathcal{E}_{SC}/\tau$ to be time independent, relying only on the rates of energy exchange, the temperature of the reservoir and the optical pump.

\section{Results}\label{sec:results}
\subsection{Quantum Battery}
\begin{figure}
  \centering
   \includegraphics[width=1.0\columnwidth]{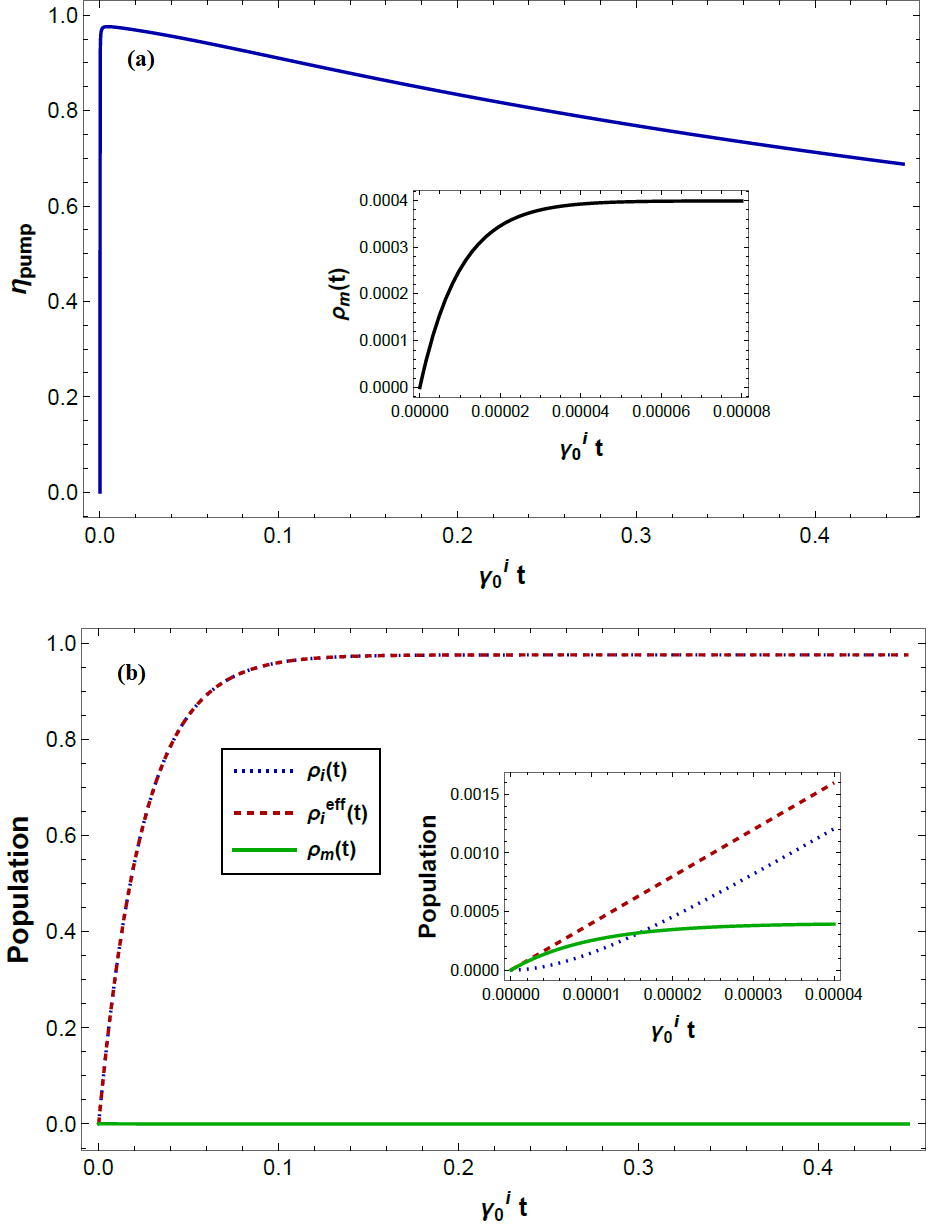}
   \caption{We plot the pumping efficiency (Fig. (a)), $\eta_{pump}$, and the population of levels $\ket{m}$ and $\ket{i}$ (Fig. (b))  as a function of $\gamma_0^{i}t$ for $T = 0$ and $\omega_f = \omega_m$. In Fig. (b), $\rho_i^{eff}$ is the population of the level $\ket{i}$ when we adiabatically eliminate level $\ket{m}$ and obtain an effective two-level system, with levels $\ket{g}$ and $\ket{i}$. Parameters: $\omega_m/\omega_i = 1.02$, $\gamma_0^m/\omega_i = 10^{-4}$, $\gamma_0^i/\omega_i = 10^{-9}$, $\Omega/\omega_i = 10^{-6}$ and $\omega_i \tau \approx 0.45 \times 10^{9}$. In the graphics, Bohr energies are given in units of $\hbar$ and thermal energies in units of $k_B$.} \label{eft}
\end{figure}

\begin{figure}
  \centering
   \includegraphics[width=1.0\columnwidth]{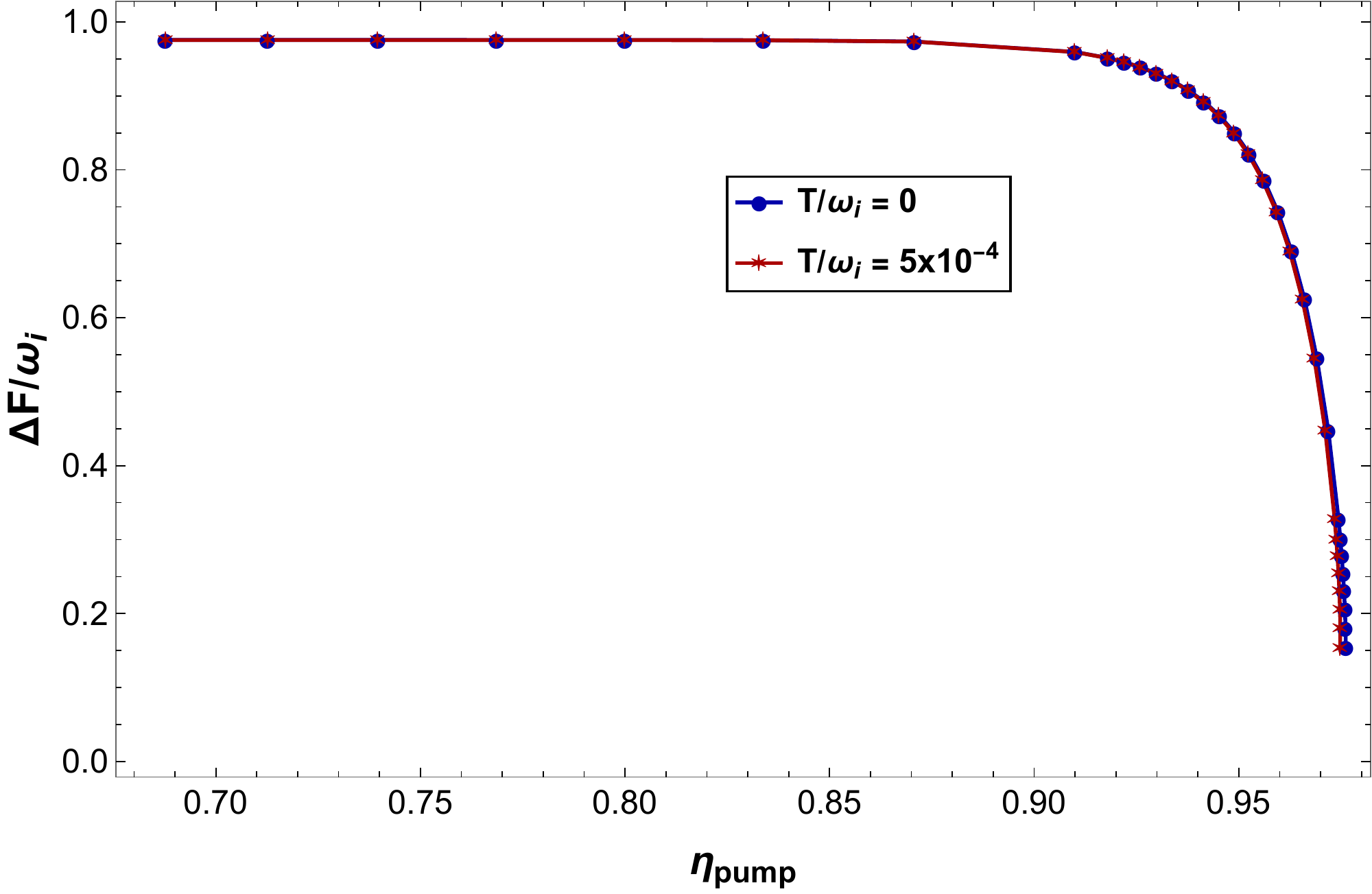}
   \caption{Stored energy versus efficiency of the charging process for two temperatures. Parameters: $\omega_m/\omega_i = 1.02$, $\omega_f = \omega_m$, $\gamma_0^m/\omega_i = 10^{-4}$, $\gamma_0^i/\omega_i = 10^{-9}$, $\Omega/\omega_i = 10^{-6}$.} \label{DeltaFxEff}
\end{figure}

The first aspect to analyze regarding the charging of the battery via optical pumping is the time scales of the process. In Figs.~\ref{eft} (a) and (b) we present respectively the numerical results for the pumping efficiency and the population of level $|i\rangle$ as a function of (parameterized) time. We consider a resonant pump ($\omega_f = \omega_m$) and parameters consistent with the adiabatic elimination of level $|m\rangle$ ($\gamma^-_m = 10^5\gamma^-_i$ and $\gamma^-_m=10^2\Omega$). In Fig.~\ref{eft} (a), we show that there are clearly two time scales involved in the process, as expected. The first one, much faster, concerns the coarse-graining in time required to take the adiabatic elimination of level $|m\rangle$ into consideration. Its characteristic time is of the order of $1/\gamma^-_m$ as it becomes clear in the inset where we show how the population $\rho_m$ quickly stabilizes in time when compared to the effective dynamics of the battery dominated by $\gamma^-_i$. At $t=0$, there is no stored power in the battery and the very first process coherently transfers population from level $|g\rangle$ to level $|m\rangle$. As the decay from $|m\rangle$ to $|i\rangle$ begins to dominate the process, the battery begins to store energy and, the closer level $|m\rangle$ gets to its asymptotic value, the more efficient this energy transference gets until the moment where almost all the energy pumped into the system is stored as population of level $|i\rangle$. That is when the efficiency reaches its peak, at which point the only wasted energy is the heat dissipated to the $|m\rangle\rightarrow |i\rangle$ reservoir and the little remaining population in level $|m\rangle$. From the efficiency point of view, this is the best operating interval for charging the battery. A glance on Fig.~\ref{eft} (b), however, shows that at this time scale, the battery is far from fully charged. In fact, the population of level $|i\rangle$ is still comparable to that of level $|m\rangle$ and much smaller than its asymptotic value, as it is made explicit by the inset of the figure. The further we pump the system, the more stored power we get at the cost of lowering the efficiency of the process because, now, the system begins to dissipate heat to the $|i\rangle\rightarrow |g\rangle$ reservoir as well. This becomes clear in Fig.\ref{DeltaFxEff} where we plot the stored energy versus the efficiency for two different temperatures of the natural reservoir. In each curve, both plotted quantities are parameterized by the duration of the charging cycle of the battery for a fixed set of the remaining parameters. Note that the temperature does not significantly affect the process except for very short times (higher efficiencies). 

We also show in Fig.~\ref{eft} (b) that the chosen parameters are indeed consistent with the adiabatic elimination condition since, at the time scale of the charging of the battery, the population of level $|i\rangle$ is essentially the same taking both the full three-level or the effective two-level dynamics.

\begin{figure}
  \centering
   \includegraphics[width=1.0\columnwidth]{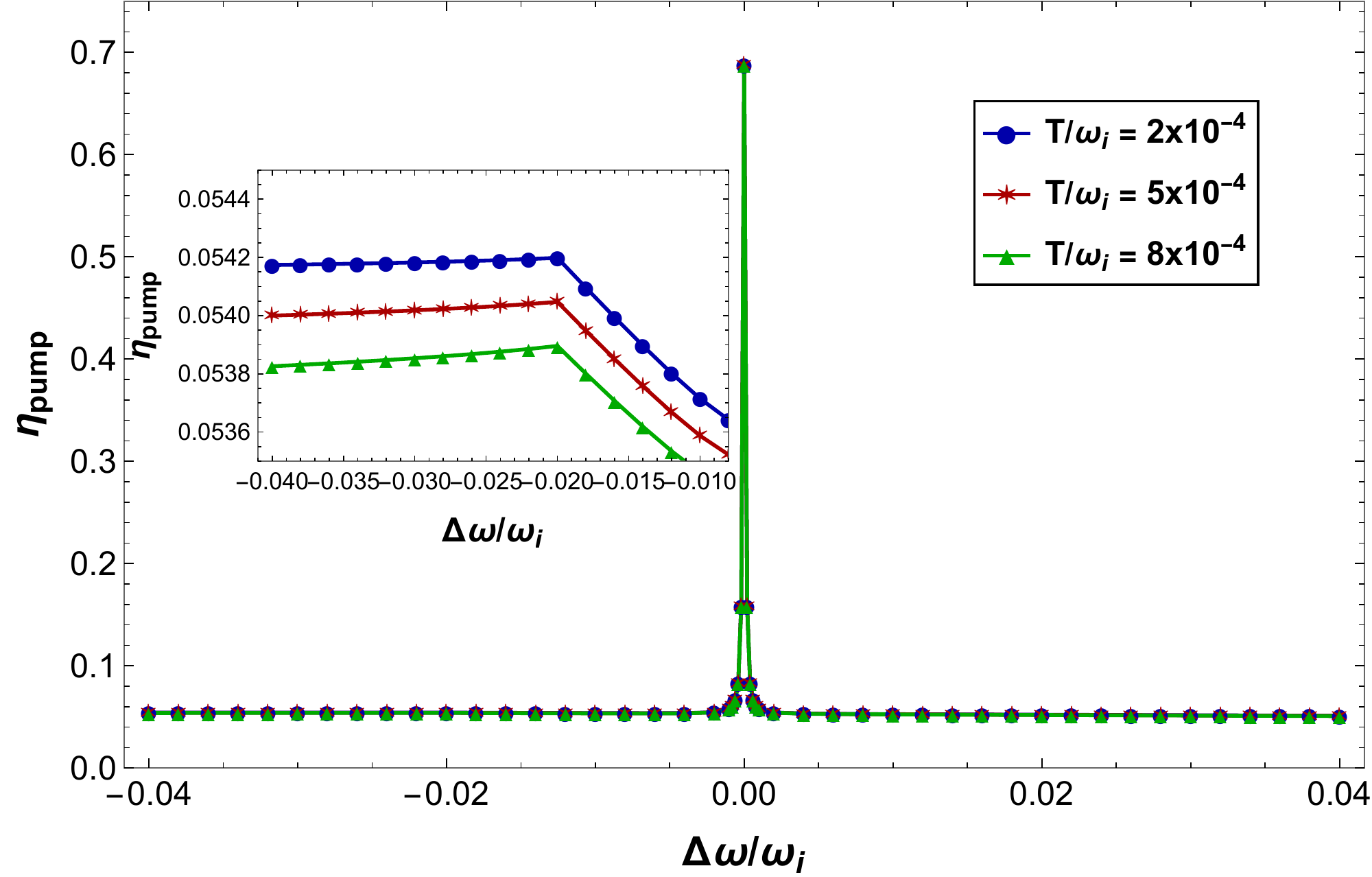}
   \caption{Pumping efficiency, $\eta_{pump}$, as a function of $\Delta \omega/\omega_i$, where $\Delta \omega = \omega_f - \omega_m$, for different temperatures. Parameters: $\Omega/\omega_i = 10^{-6}$, $\gamma_0^m/\omega_i = 10^{-4}$, $\gamma_0^i/\omega_i = 10^{-9}$ and $\omega_m/\omega_i = 1.02$.} 
   \label{efp}
\end{figure}

The next relevant point concerning the battery is how its charging efficiency depends on the frequency of the pumping field. In Fig.\ref{efp} we plot the efficiency of the optical pumping process as a function of the difference in energy $\Delta \omega = \omega_f-\omega_m$ between the photon coming from the external drive, of frequency $\omega_f$, and the $|g\rangle \rightarrow |m\rangle$ transition, of frequency $\omega_m$. First of all, note the existence of a resonance peak when the frequency of the pump field matches the energy gap of level $|m\rangle$ ($\omega_f = \omega_m$). The height of this peak increases with the pumping rate $p$, saturating for values of $p$ that maximize the population of level $|i\rangle$. At this point, the only main source of efficiency degrading is the already mentioned heat wasted in the $|m\rangle \rightarrow |i\rangle$ channel. This heat is proportional to the difference $E_m-E_i$ and will increase (lowering the efficiency) for higher energy gaps. 

\begin{figure}
  \centering
   \includegraphics[width=1.0\columnwidth]{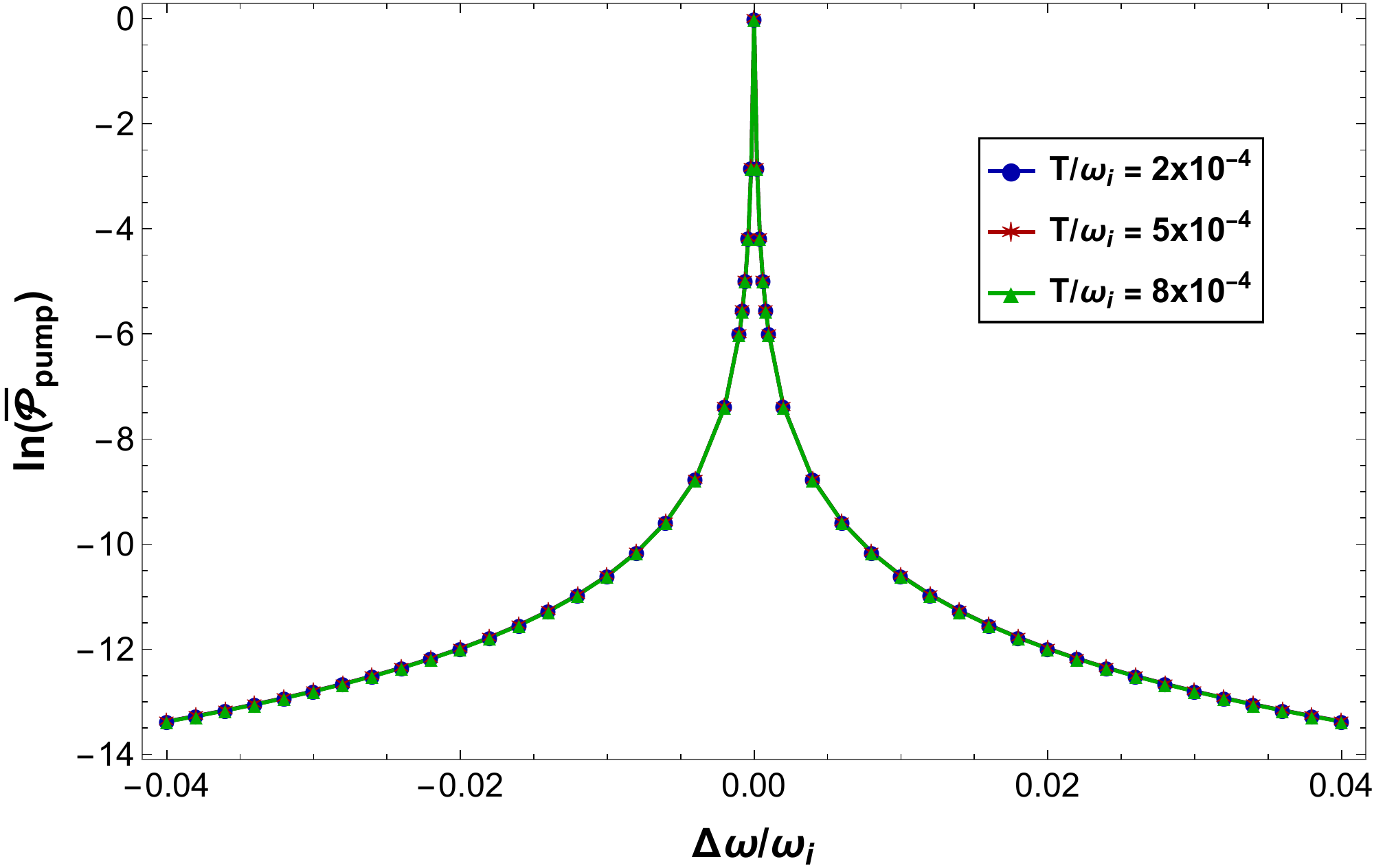}
   \caption{We plot the logarithm of the normalized input power, $\bar{\mathcal{P}}_{pump} = \mathcal{P}_{pump}/\mathcal{P}_{pump}^{max}$, where $\mathcal{P}_{pump}$ is the input power and $\mathcal{P}_{pump}^{max}$ is the maximum input power obtained at $\Delta \omega = 0$   $(\omega_f = \omega_m)$, as a function of $\Delta \omega/\omega_i$ for different temperatures. We use the same set of parameters as in figure \eqref{efp}.} \label{powerp}
\end{figure}

When pumping off-resonantly, the detuning $\Delta \omega = \omega_f-\omega_m$ is one of the main factors that affect the efficiency. This can be seen in the inset of Fig.~\ref{efp}: there is a decay of the efficiency as soon as $\omega_f >\omega_i$ and until resonance is reached. The steady efficiency for lower pumping frequencies ($\omega_{f} \leq \omega_i$) is due to a compensation mechanism in which the heat flux from $|m\rangle$ to $|i\rangle$ inverts signal and becomes positive, injecting energy into the system to complement the energy coming from the work done by the outside source. Note, however, that the very low pump power at this range (displayed in Fig.~\ref{powerp}) generates really low efficiencies and stored energies in the battery, making the resonance condition by far the best one to transfer energy into it.

The other mechanism that affects the efficiency is the energy fluctuation due to the natural reservoir. Its first and clearest effect comes from the fact that, for fixed values of the remaining parameters, the lower the temperature of the reservoir the higher the efficiency of the optical pumping. As we have seen, in the limit of very low $T$, these fluctuations do not play a significant role and the only energy waste comes from the mechanisms described above. As $T$ increases, the relative effect of the pumping becomes less pronounced and the efficiency of the process decreases.

\begin{figure}
  \centering
   \includegraphics[width=1.0\columnwidth]{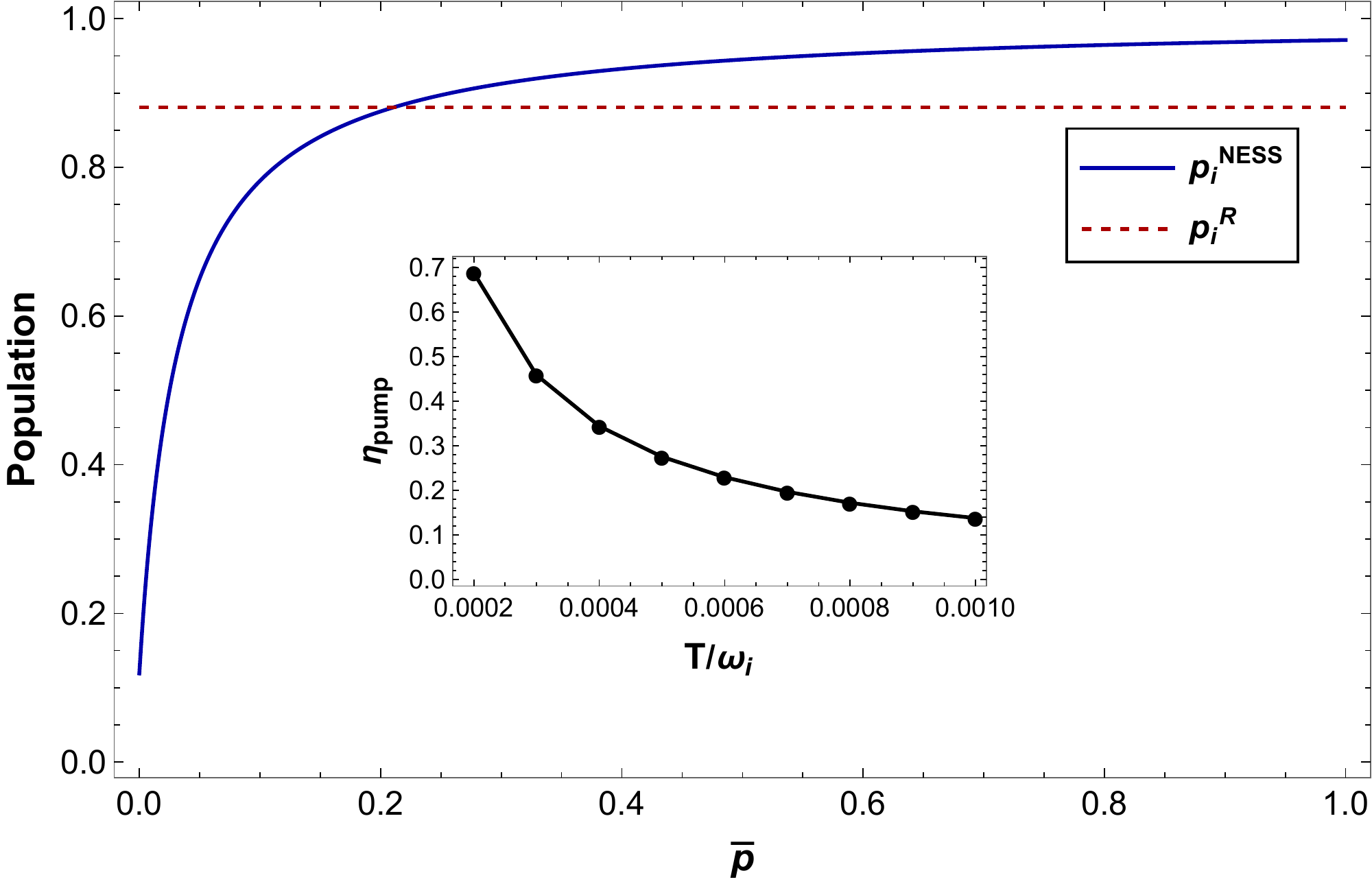}
   \caption{We plot the population of level $\ket{i}$ for a Rabi flip, $p_i^R$, and for the optical pumping in the steady state, $p_i^{NESS}$ as a function of $\bar{p} = p/p_{max}$. We vary $\Omega$ from $10^{-9} \omega_i$ to $10^{-6} \omega_i$ and we kept the other parameters fixed. To calculate $p_{max}$ we use $\Omega/\omega_i = 10^{-6}$. Parameters: $T/\omega_i = 0.5$, $\omega_f = \omega_m = 5 \omega_i$, $\gamma_0^m/\omega_i = 10^{-4}$ and $\gamma_0^i/\omega_i = 10^{-9}$. In the inset we plot the efficiency as a function of the temperature for a fixed ratio $K_B T/\hbar\omega_m$ and for $\omega_f = \omega_m$. Inset parameters: $\Omega/\omega_i = 10^{-6}$, $\gamma_0^m/\omega_i = 10^{-4}$ and $\gamma_0^i/\omega_i = 10^{-9}$} \label{popp}
\end{figure}

Note, however, that, as long as the conditions for the adiabatic elimination of level $|m\rangle$ hold, meaning as long as the temperature is not too high for its thermal population to become significant, the efficiency becomes approximately independent of the temperature of the reservoir at the resonance. Therefore, temperature itself is not the main issue regarding the charging of the battery. As mentioned before, as long as $K_BT \ll E_m-E_i$, one can still eliminate level $|m\rangle$ and fully charge the battery, even for high temperatures. For example, in fig.~\ref{popp} we plot the power stored in the battery as a function of the pumping power $p$ for a much higher temperature ($K_B T/\hbar \omega_i \equiv 0.5$). In order to guarantee the validity of our protocol, we have also raised the energy gap of level $|m\rangle$ from $E_m/E_i = 1.02$ to $E_m/E_i = 5$. Note that for high enough values of $p$, we can still fully charge the battery. Just for the sake of completeness, we have also included in Fig.~\ref{popp} a line representing the best charging of the battery that a unitary (hence isentropic) protocol could achieve at the same temperature. This unitary can be, for example, a Rabi flip induced by directly coupling levels $|g\rangle$ and $|i\rangle$ and it has efficiency one. However, this method of charging the battery is limited by the entropy of its initial state: at best, it inverts the initial populations of level $|g\rangle$ and $|i\rangle$, a limitation not shared by optical pumping which can produce $p_i \approx 1$. This difference becomes more and more accentuated as the temperature of the environment surrounding the battery increases. The price to pay, on the other hand, is the downgrade of the efficiency of the optical pumping: higher temperatures require higher energy separations $E_m-E_i$ which lead to a larger heat dissipation due
to the $\ket{m} \rightarrow \ket{i}$ decay. In the inset of Fig.~\ref{popp} we plot the efficiency as a function of the temperature for maximized pumping rate $p$ and a fixed ratio $K_B T/\hbar \omega_m$ (i.e. fixed initial population of level $|m\rangle$).
\subsection{Quantum heat engine}
We start by analyzing the performance of the two-stroke machine in the short cycle limit ($\sum_j \gamma_j \tau \ll 1$) which was previously proven to be the most efficient one. In this scenario, the efficiency and the output power are given by $\eta_{SC} = \mathcal{E}_{SC}/E_{in}^{SC}$ and $\mathcal{P}_{SC} = \mathcal{E}_{SC}/\tau$, respectively, where $\mathcal{E}_{SC}$ is given by \eqref{EFFSC} and $E_{in}^{SC}$ is calculated similarly to what was done for the quantum battery in section~\ref{QBM}

\begin{figure}
  \centering
   \includegraphics[width=1.0\columnwidth]{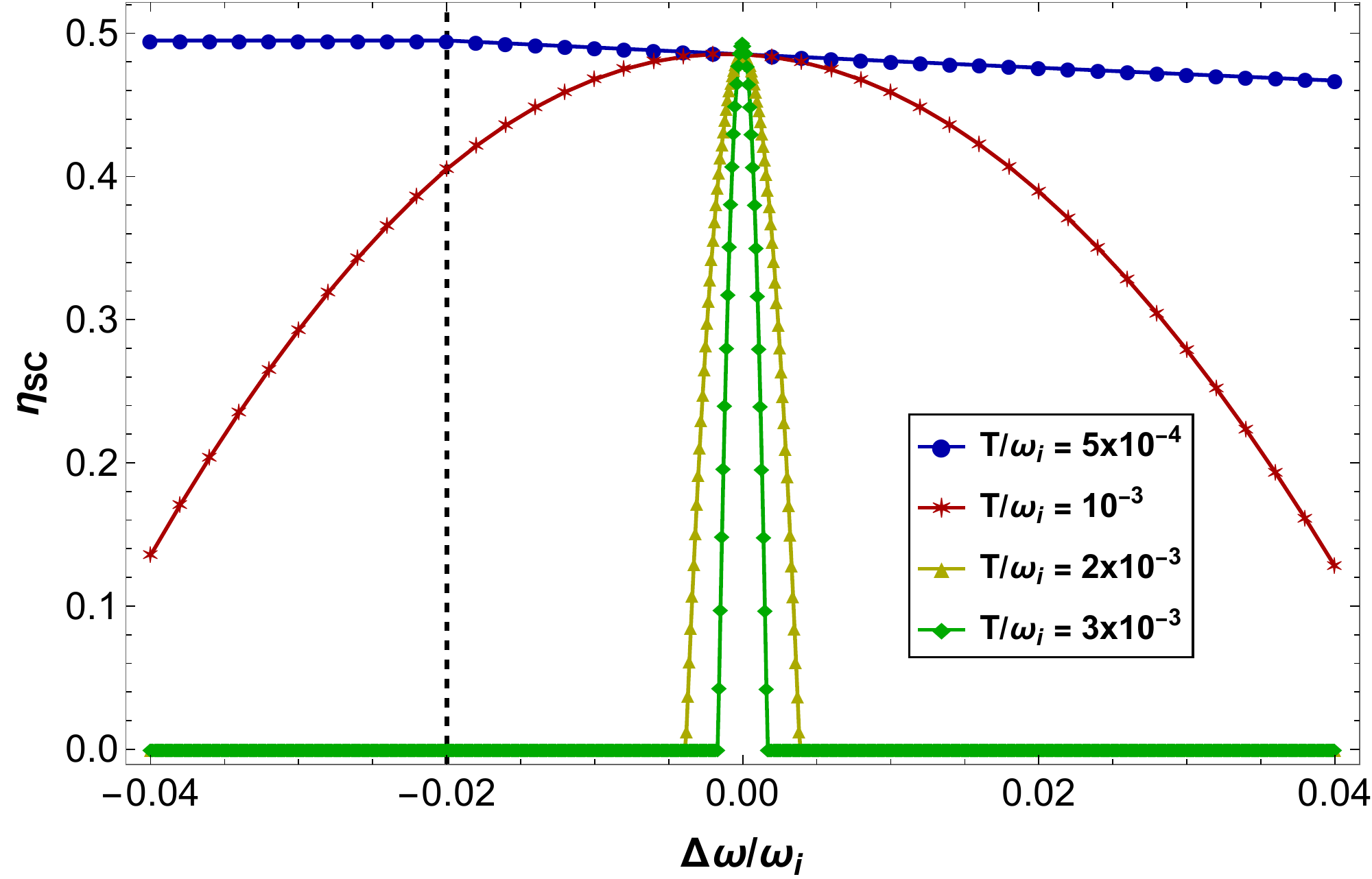}
   \caption{We plot the efficiency of the machine as a function of $\Delta \omega/\omega_i$ in the short cycle (SC) limit for different temperatures. Parameters: $\Omega/\omega_i = 10^{-6}$, $\gamma_0^m/\omega_i = 10^{-4}$, $\gamma_0^i/\omega_i = \gamma_0^e/\omega_i = 10^{-9}$, $\omega_e/\omega_i =0.01$ and $\omega_m/\omega_i = 1.02$.} \label{efsc}
\end{figure}

\begin{figure}
  \centering
   \includegraphics[width=1.0\columnwidth]{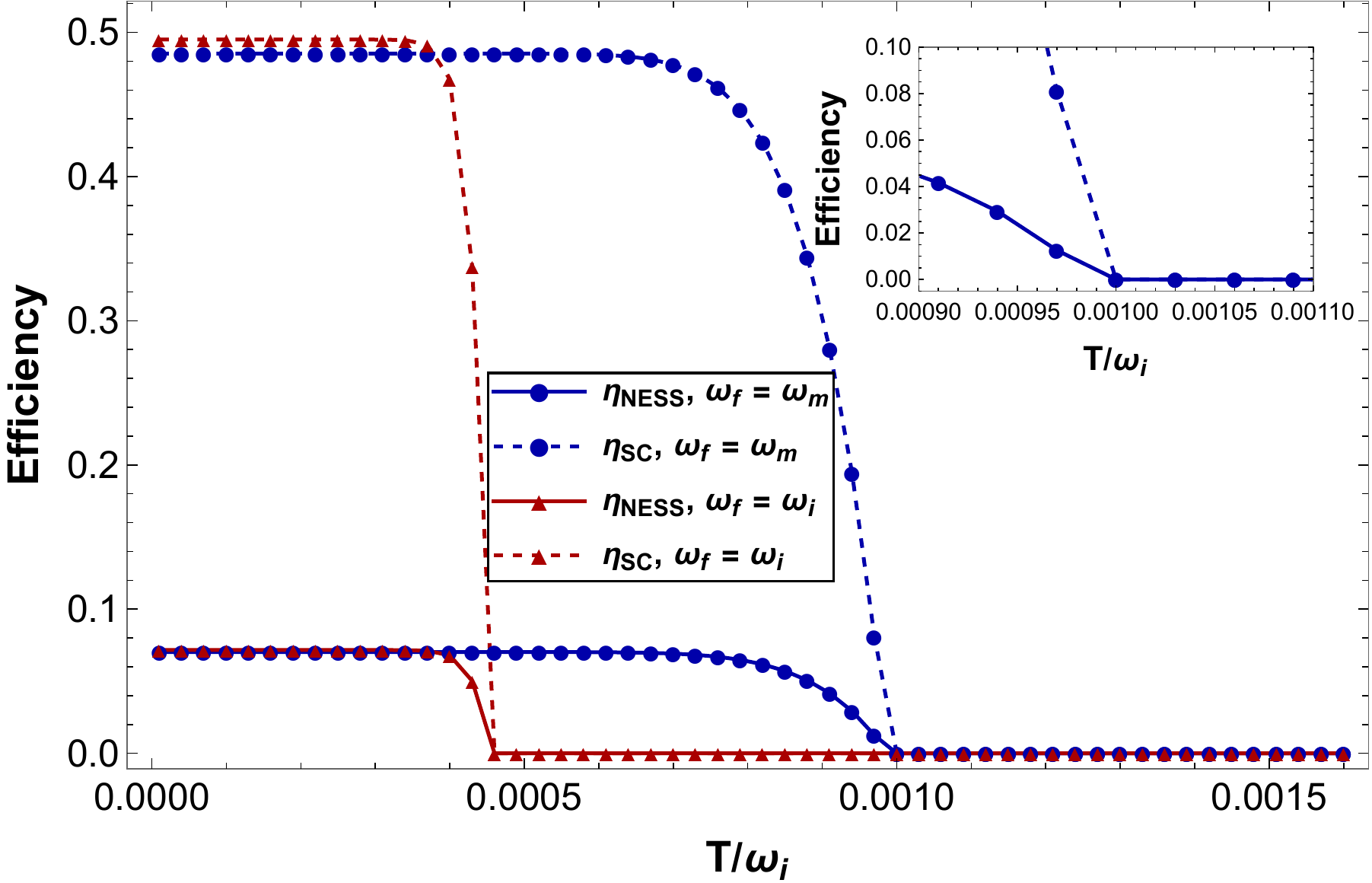}
   \caption{Here we plot the efficiency in the asymptotic cycle, $\eta_{NESS}$ and the efficiency in the short cycle, $\eta_{SC}$, for different values of the frequency of the external drive, $\omega_f$, as a function of the temperature of the natural thermal reservoir. Parameters: $\Omega/\omega_i = 10^{-8}$, $\gamma_0^m/\omega_i = 10^{-4}$, $\gamma_0^i/\omega_i = \gamma_0^e/\omega_i = 10^{-7}$, $\omega_e/\omega_i =0.01$, $\omega_m/\omega_i = 1.02$ and $\epsilon/\omega_i = 2 \times 10^{-4}$ (for the asymptotic cycle).} \label{eflong}
\end{figure}

In fig. \ref{efsc}, we plot the efficiency as a function of $\Delta \omega$. Note that, except for very low temperatures, $\eta_{SC}$ is maximized at resonance, $\omega_f = \omega_m$. Again, this happens because that is when the pumping rate, $p$, reaches its maximum value, injecting the most possible energy into the work fluid to be later extracted in the discharging stage. Contrary to the battery, however, the machine does not work at any temperature and for any detuning. Here, it is not enough to take the system out of thermal equilibrium by storing some energy in it. By design, the machine requires positive ergotropy after the recharging stage, which is achieved for $\rho_{ii} > \rho_{ee}$ and this condition cannot be matched if $T$ is too high and/or $p$ is too low. In fact, as the temperature of the natural reservoir increases, so does the thermal population of level $|e\rangle$. This requires more pumping power to invert the population in the $\{|e\rangle,|i\rangle\}$ subspace limiting the operation of the machine to detunings around the resonance. The higher the temperature, the closer to resonance one needs to pump. In fact, as we see in Fig.~\ref{eflong}, there is a temperature threshold beyond which the machine does not work, no matter how strongly we pump. We comment more on this later in this section. 

Thermal fluctuations also do affect the overall efficiency of the machine in the short cycle but in a much smaller scale than its range of operation, i.e. as long as the machine works, its efficiency is close to the maximum for a given pumping frequency.

As we have seen in section~\ref{QHE}, the efficiency of the short cycle is maximized when $\frac{\Gamma^+_i}{\gamma_i^-} \gg \frac{\gamma^+_e}{\gamma_e^-}$. If we consider a standard thermal reservoir, then one of the ways to achieve this limit is to operate coupled to a very low temperature ($T \ll 1$) and in resonance. In this limit, $\Gamma^+_i \sim p\gamma^-_m$, $\gamma^+_e \approx 0$ and the efficiency of the short cycle is approximated by 
\begin{equation}
    \eta_{SC} \approx \frac{\gamma_e^-}{\gamma_e^- + \gamma_i^-}\frac{\omega_i - \omega_e}{ \omega_m}=\frac{\gamma_e^-}{\gamma_e^- + \gamma_i^-}\frac{\omega_i - \omega_e}{ \omega_i}\frac{1}{1+\frac{\delta \omega}{\omega_i}},
\end{equation}
where $\delta\omega = \omega_m-\omega_i$. This result approaches the efficiency of the Otto cycle operating in a three-level scheme, $\eta_{Otto}=\frac{\omega_i - \omega_e}{ \omega_i}$, when $\gamma_e^- \gg \gamma_i^-$. $\eta_{Otto}$ was shown to be the efficiency of this two-stroke machine operating under thermal reservoirs of temperatures $T_H$ and $T$ affecting respectively the $|g\rangle\rightarrow |i\rangle$ and the $|g\rangle\rightarrow |e\rangle$ transitions~\cite{Tiago2021}. The extra factor $\frac{1}{1+\frac{\delta \omega}{\omega_i}}$, that reduces this efficiency, comes from the energy wasted to generate the effective temperature $T_H$ through optically pumping level $|m\rangle$. Note that, as long as $\delta \omega \ll \omega_i$, this correction amounts to a small linear decay of the efficiency proportional to $\frac{\delta \omega}{\omega_i}$. This factor, however, cannot be made as small as possible because there must be a minimum energy separation between levels $|i\rangle$ and $|m\rangle$ to make it possible to adiabatically eliminate the latter.

\begin{figure}
  \centering
   \includegraphics[width=1.0\columnwidth]{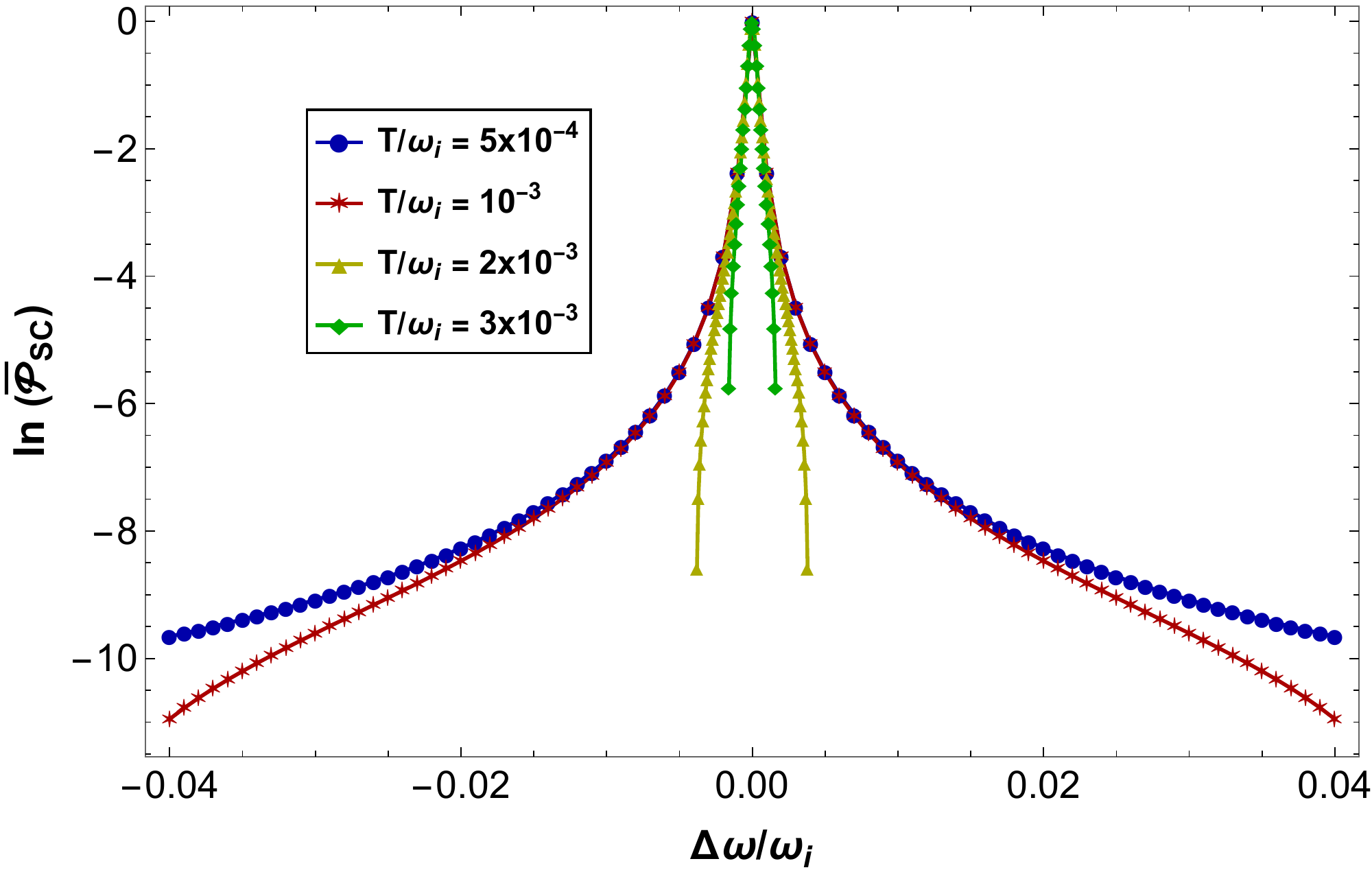}
   \caption{We plot the logarithm of the normalized output power, $\bar{\mathcal{P}}_{SC} = \mathcal{P}_{SC}/\mathcal{P}_{SC}^{max}$, where $\mathcal{P}_{SC}$ is the output power and $\mathcal{P}_{SC}^{max}$ is the maximum output power, as a function of $\Delta \omega/\omega_i$ for different temperatures. We use the same parameters as in fig. \ref{efsc}.} \label{psc}
\end{figure}

The other relevant figure of merit of the heat machine is its output power. In fig. \ref{psc} we see that this quantity reaches its maximum value when the external source drives the $|g\rangle \rightarrow |m\rangle$ transition resonantly, $\omega_f = \omega_m$, and decays symmetrically around it. We also note that around resonance, $\omega_f \sim \omega_m$, as long as the positive ergotropy condition is achieved, the output power does not depend significantly on the temperature of the natural reservoir. However, as previously analysed, the higher the temperature, the closer to resonance one needs to be for the machine to operate. 

From the efficiency and power analysis, we can state that, as in the quantum battery, the machine works better in the regime of low temperature. It is worth noting that the short cycle gives the best performance but the output power per cycle, that is proportional to the stored ergotropy, is very small. That is the downside of maximizing the efficiency.

We can also analyse the performance of the machine operating in the limit of the asymptotic cycle, i.e. when the duration of the recharging stage, $\tau_r$ is set to allow the battery to store close to its maximum capacity. In this limit, the operational steady state, $\rho_{OSS}$, converges to $\rho_{NESS}$ and the efficiency and the output power are given by $\eta = - W_{ext}/E_{in}$ and $\mathcal{P} = - W_{ext}/\tau$, respectively, where $W_{ext}$ is calculated via Eq. \eqref{alickiwork}. In fig. \ref{eflong} we plot the efficiency of both the short and long cycles as a function of the temperature for two pumping frequencies. Note that, for a given detuning, the temperature threshold is the same for both cycles. The largest threshold, on the other hand, is found at resonance ($\omega_f = \omega_m$) whereas, for fixed couplings to the external drive, larger detunings lower the temperature threshold, as expected.

The dependence of the output power with the duration of the cycle is the same as the one observed in previous works~\cite{Tacchino2020,Tiago2021}: the power is maximized for the short cycle and decays as the cycle duration increases. Considering that the output power is the ratio between the extracted work and the duration of the cycle $\tau$, that means that the increase in the extraction of work grows slower than linearly with $\tau$, the linear increase being the one obtained exactly in the short cycle.

\section{conclusion} \label{sec:conclusion}
This work have encompassed two studies: first, we have studied the optical pumping of a quantum battery from the perspective of the efficiency, the input power and the total energy stored in the process. We have concentrated our analysis in a cascade three-level scheme where the upper energy level is adiabatically eliminated from the dynamics. In this particular scenario, we have found that there is no universal optimal set of parameters to charge the battery. The process is more efficient for short times, in a time scale comparable with the adiabatic elimination condition, dominated by the decay rate of the accessory higher energy level $|m\rangle$ ($t\sim 1/\gamma^-_m$). This is consistent with the idea that the stabilization of the population of the eliminated level marks the moment in time when most of the pumped energy is directly transferred to the optically pumped target level of the battery. This is the best achievable efficiency in the process but, in this time scale, the battery is still quite empty. The other meaningful time scale is the one to the fully charge the battery, connected to the decay time of the optically pumped level ($t\sim 1/\gamma^-_i$). At this time scale, the battery stores as much power as possible but the overall efficiency drops because some of the pumped energy is lost in the form of heat. We have also shown that, as long as the adiabatic elimination is still valid, i.e. as long as the temperature of the dissipative thermal reservoir is not too high compared to the energy gap of the eliminated level, temperature does not affect significantly the charging process. Finally, the charging process is much more efficient in resonance ($\omega_m =\omega_f$), justifying this condition as the best charging protocol.

The second part of the work regards using the battery as the working fluid of a two-stroke thermal machine. Once again, best performance is achieved operating at low temperature and in short cycles. The maximum efficiency approaches that of an equivalent machine running an Otto cycle between two different reservoirs. The correcting factor that reduces the Otto cycle efficiency comes from the energy wasted to create the effective high temperature reservoir through the optical pumping mechanism and can be made small as long as the adiabatically eliminated level is not much more energetic than the optically pumped one. The Otto cycle efficiency cannot be achieved, though, because for the adiabatic elimination to work, some energy separation is still required between the levels.

Still concerning the short cycle, highest efficiency is achieved when the pumping frequency matches the Bohr frequency of the optically pumped transition ($\omega_f=\omega_i$). However, differently from the first part of the work, the analysed two-stroke machine cannot operate at any combination of pump frequency and temperature. The higher the cold reservoir temperature, the closer to resonance ($\omega_f \sim \omega_m$) the machine needs to be. On the other hand, in resonance and as long as the machine is able to operate, its efficiency is not significantly affected by the temperature of the cold reservoir. The shutdown transition due to the temperature raise is quite sharp.

We also analysed the machine under the asymptotic cycle, where the battery is allowed to be fully charged in the recharging state. We have shown that it has the same shutdown temperature of the short cycle, i.e. the threshold temperature of the cold reservoir beyond which work extraction stops is the same regardless of the duration of the cycle. Efficiency and output power are both lower in this regime, however, as it has already been shown in previous works~\cite{Tacchino2020,Tiago2021}.

Finally, it is worth mentioning that the models adopted here can be adapted to different quantum optical setups. Both the quantum battery and the heat engine can be implemented in both natural and artificial atoms, superconducting circuits and similar systems.
\section{Acknowledgments}

M.F.S. acknowledges FAPERJ Project No. E-26/202.576/2019 and CNPq Projects No. 302872/2019-1 and INCT-IQ 465469/2014-0. T.F.F.S. acknowledges CAPES for financial support.

\section{Appendix} 

Here we show more details about the calculations of the results shown in the main text.

\subsection{Quantum Batterry}

\subsubsection{Adiabatic Elimination}
The dynamics of the system is described by a master equation in the Lindblad form ($\hbar = 1$ and $k_B = 1$):
\begin{equation}\label{me3b}
\dot{\rho}(t) = -i[H(t),\rho(t)] + \mathcal{L}[\rho(t)].
\end{equation}
The Hamiltonian reads $H(t) = H_0 + V(t)$, where $H_0$ is the free Hamiltonian of the system and $V(t)$ account for coupling with an external work source. For the quantum battery model, we have $V(t) = V_{in}(t)$, where $V_{in}(t)$ is given by
\begin{equation}\label{vinapp}
    V_{in}(t) = \Omega \left( \ket{g}\bra{m} e^{i \omega_f t} + \ket{m}\bra{g} e^{-i \omega_f t} \right).
\end{equation}
The non-unitary part of the dynamics, represented by $\mathcal{L}[\rho(t)]$, is given by
\begin{equation}\label{nuappqe}
    \mathcal{L}[\rho(t)] = \mathcal{L}_{\gamma_{m}}[\rho(t)] + \mathcal{L}_{\gamma_{i}}[\rho(t)],
\end{equation}
where $\mathcal{L}_{\gamma_{m}}[\rho(t)]$ and $\mathcal{L}_{\gamma_{i}}[\rho(t)]$ are given by Eqs. \eqref{lgm} and \eqref{lgi}, respectively.

In a rotating framework, defined by $\varrho(t) = e^{i H_0^{'}t} \rho e^{-i H_0^{'}t}$, where $H_0^{'} = H_0 + \Delta \omega \ket{m}\bra{m}$ and $\Delta \omega = \omega_f - \omega_m$, Eq.~\ref{me3b} reads
\begin{equation}\label{vrho}
\dot{\varrho}(t) = -i[\bar{V},\varrho(t)] + \mathcal{L}[\varrho(t)],
\end{equation}
where
\begin{align}
    \bar{V}_{in} = \Omega (\ket{g}\bra{m} + \ket{m}\bra{g}) - \Delta\omega \ket{m}\bra{m}.
\end{align}
and $\mathcal{L}[\varrho(t)]$, is the same as in Eq.~\eqref{nuappqe}.

If $\gamma_m^{-} \gg \gamma_m^{+}, \gamma_{i}^{\pm}, \Omega$, we can adiabatically eliminate level $\ket{m}$ and its coherences ($\{\rho_{mi},\rho_{im}\}$). In particular, we obtain
\begin{align}\label{gammammb}
    \varrho_{mm} \approx \frac{1}{\gamma_{m}^{-}} [p \gamma_m^{-} \varrho_{gg} + \gamma_m^{+} \varrho_{ii}],
\end{align}
where $p = \frac{4 \Omega^2}{\gamma_{m}^{-^2} + 4 \Delta\omega^2} \ll 1$, and 
\begin{equation}\label{mggm2b}
    \varrho_{mg} - \varrho_{gm} \approx -\frac{4 i \Omega  }{\gamma_{m}^{-^2} + 4 \Delta\omega^2}\gamma_{m}^{-}(\varrho_{gg} - \varrho_{mm}).
\end{equation}
for the time evolution of the variables related to level $|m\rangle$. 

Substituting Eq. \eqref{gammammb} in the equation of motion for level $\ket{i}$, we end up with
\begin{equation}\label{iib2}
    \dot{\varrho}_{ii}(t) = \Gamma_i^{+} \varrho_{gg} - \gamma_i^{-} \varrho_{ii},
\end{equation}
where $\Gamma_i^{+} = \gamma_i^{+} + p \gamma_m^{-}$. For level $\ket{g}$ and the coherence between level $\ket{g}$ and $\ket{i}$, we obtain
\begin{equation}\label{ggb2}
     \dot{\varrho}_{gg}(t) = \gamma_i^{-} \varrho_{ii} - \Gamma_i^{+} \varrho_{gg},
\end{equation}
and
\begin{equation}\label{gib2}
    \dot{\varrho}_{gi}(t) = - i p \Delta\omega \varrho_{gi} - (\Gamma_i^{+} + \gamma_i^{-}) \varrho_{gi}.
\end{equation}
Using these equations we can, finally, write a master equation for the effective qutrit in the Lindblad form
\begin{align}
    \dot{\varrho}(t) &= - i [\bar{V}_{eff}, \varrho(t)] \nonumber \\
    &+ \frac{\Gamma_i^{+}}{2}(2 \ket{i}\bra{g} \varrho(t) \ket{g} \bra{i} - \{\ket{g}\bra{g}, \varrho(t)\}) \nonumber \\
    &+ \frac{\gamma_i^{-}}{2}(2 \ket{g}\bra{i} \varrho(t) \ket{i} \bra{g} - \{\ket{i}\bra{i}, \varrho(t)\})
\end{align}
where $\bar{V}_{eff} = p \Delta\omega \ket{g}\bra{g}$ and $\{A, B\} = AB + BA$. This equation are the same as the ones for a qubit in contact with a thermal reservoir at temperature $T_H=\frac{\hbar\omega_{i}}{k_B\textrm{ln}(\frac{\gamma^-_i}{p\gamma^-_m+\gamma_i^+})}$.

\subsubsection{Thermodynamics}

The efficiency of the pumping process, $\eta_{pump}$, and its input power,$\mathcal{P}_{pump}$, are defined respectively by
\begin{equation}
    \eta_{pump} = \frac{\Delta F}{E_{in}}
\end{equation}
and
\begin{equation}
    \mathcal{P}_{pump} = \frac{\Delta F}{\tau},
\end{equation}
where $\Delta F$ is the variation of the Helmholtz free energy during the process, from $t_0 = 0$ to $t=\tau$. and $E_{in}$ is the energy injected into the system

\begin{equation}\label{einb}
    E_{in} = Max\{W_{in}, W_{in} + Q_{\gamma_m}, W_{in} + Q_{\gamma_i}, W_{in} + Q\}.
\end{equation}
Here $W_{in}$ is the work done on the system, given by
\begin{align}\label{w3bn}
  W_{in} &= \int_{0}^{\tau} dt \operatorname{Tr}\{\rho(t) \dot{V}_{in}(t)\} \nonumber \\
     &=\int_{t_0}^{t_f} dt \operatorname{Tr} \{\rho^j[ i \Omega \omega_f (\ket{g}\bra{m} e^{i \omega_f t} - \ket{m}\bra{g} e^{-i \omega_f t})]\} \nonumber \\
      &=\int_{t_0}^{t_f} dt \hskip 0.1cm i \Omega \omega_f [\rho_{mg}(t)e^{i \omega_f t} - \rho_{gm}(t)e^{-i \omega_f t}] \nonumber \\
      &= i \Omega \omega_f \int_0^{\tau} dt \hskip 0.1cm [\varrho_{mg}(t) - \varrho_{gm}(t)],
\end{align}
where $\varrho = e^{i H_0^{'}t} \rho e^{-i H_0^{'}t}$.  $Q_{\gamma_m}$ is the heat exchange associated with the transitions between levels $\ket{i}$ and $\ket{m}$,
\begin{align}\label{qm3bn}
   Q_{\gamma_m} &=  \int_0^{\tau} dt \operatorname{Tr}[\mathcal{L}_{\gamma_m}[\rho(t)]H(t)] \nonumber \\
   &= -\Omega \frac{\gamma_m^{-}}{2} \int_0^{\tau} dt \hskip 0.1cm [\varrho_{mg}(t) + \varrho_{gm}(t)] \nonumber \\
   &+ (\omega_{m} - \omega_{i}) \int_0^{\tau} dt \hskip 0.1cm [\gamma_m^{+} \varrho_{ii}(t) - \gamma_m^{-} \varrho_{mm}(t)],
\end{align}
$Q_{\gamma_i}$ with the transitions between levels $\ket{g}$ and $\ket{i}$,
\begin{align}\label{qi3bn}
   Q_{\gamma_i} &=  \int_0^{\tau} dt \operatorname{Tr}[\mathcal{L}_{\gamma_i}[\rho(t)]H(t)] \nonumber \\
   &= -\Omega \frac{\gamma_i^{+}}{2} \int_0^{\tau} dt \hskip 0.1cm [\varrho_{mg}(t) + \varrho_{gm}(t)] \nonumber \\
   &+ \omega_i \int_0^{\tau} dt \hskip 0.1cm [\gamma_i^{+} \varrho_{gg}(t) - \gamma_i^{-} \varrho_{ii}(t)]
\end{align}
and $Q =  Q_{\gamma_m} + Q_{\gamma_i}$ is the total heat exchange in the system.

In the adiabatic elimination regime, substituting Eq. \eqref{mggm2b} in Eq. \eqref{w3bn}, we obtain
\begin{equation}
    W_{in} = p \gamma_m^{-} \omega_f \int_0^{\tau} dt \hskip 0.1cm (\varrho_{gg}(t) - \varrho_{mm}(t)).
\end{equation}
In this limit, $\varrho_{gg} \gg \varrho_{mm}$, and
\begin{equation}
    W_{in} \approx p \gamma_m^{-} \omega_f \int_0^{\tau} dt \hskip 0.1cm \varrho_{gg}(t).
\end{equation}

We can make similar approximations for the calculation of the heat exchange. The sum of the coherences is approximately given by
\begin{equation}\label{gmmg3b}
    \varrho_{mg}(t) + \varrho_{gm}(t) \approx  \frac{8 \Omega}{\gamma_m^{-^{2}} + 4 \Delta \omega^2} \Delta \omega (\varrho_{gg}(t) - \varrho_{mm}(t)),
\end{equation}
and substituting this expression and Eq. \eqref{gammammb} in Eqs. \eqref{qm3bn} and \eqref{qi3bn}, we obtain
\begin{align}
     Q_{\gamma_m} &\approx - p \gamma_{m}^{-} \Delta \omega \int_0^{\tau} dt \hskip 0.1cm \varrho_{gg}(t) \nonumber \\
     &- (\omega_m - \omega_i) p \gamma_{m}^{-} \int_0^{\tau} dt \hskip 0.1cm \varrho_{gg}(t) \nonumber \\
     &= p \gamma_{m}^{-} (\omega_i - \omega_f) \int_0^{\tau} dt \hskip 0.1cm \varrho_{gg}(t)
\end{align}
and
\begin{align}
     Q_{\gamma_i} = &- p \gamma_{i}^{+} \Delta \omega \int_0^{\tau} dt \hskip 0.1cm \varrho_{gg}(t) \nonumber \\
     &+\omega_i \int_0^{\tau} dt \hskip 0.1cm (\gamma_i^{+} \varrho_{gg}(t) - \gamma_i^{-} \varrho_{ii}(t)).
\end{align}
Once again, we consider that $\varrho_{gg} \gg \varrho_{mm}$.

\subsection{Quantum heat engine}

\subsubsection{Adiabatic Elimination}

The calculations for the adiabatic elimination of the level $\ket{m}$ for the two stroke quantum heat engine are similar to those of the quantum battery. The dynamics of the system is also described by a master equation in the Lindblad form, given by Eq. \eqref{me3b}. The quantum heat engine includes a fourth level, $\ket{e}$, so the non-unitary part of the dynamics has an extra term and given by
\begin{equation}\label{nuapp}
    \mathcal{L}[\rho(t)] = \mathcal{L}_{\gamma_{m}}[\rho(t)] + \mathcal{L}_{\gamma_{i}}[\rho(t)] + \mathcal{L}_{\gamma_{e}}[\rho(t)],
\end{equation}
where $\mathcal{L}_{\gamma_{m}}[\rho(t)]$ and $\mathcal{L}_{\gamma_{i}}[\rho(t)]$ are given by Eqs. \eqref{lgm} and \eqref{lgi}, respectively, and $\mathcal{L}_{\gamma_{e}}[\rho(t)]$ is given by \eqref{lge}.

For the unitary part of the dynamics we have that in the discharging stage $V(t)$ has an extra term, $V_{ext}$, that extracts the energy stored in the system in the form of work and is given by
\begin{equation}\label{vext4qe}
    V_{ext}(t) = \epsilon \left( \ket{e}\bra{i} e^{i (\omega_i - \omega_e) t} + \ket{i}\bra{e} e^{-i (\omega_i - \omega_e) t} \right).
\end{equation}
In the recharging stage we have that $V(t) = V_{in}(t)$, as in the quantum battery, given by Eq. \eqref{vinapp}.

In the same rotating framework used before, the master equation for the engine becomes
\begin{equation}
\dot{\varrho}(t) = -i[\bar{V},\varrho(t)] + \mathcal{L}(\varrho(t)),
\end{equation}
where in the discharging stage we have 
\begin{align}
    \bar{V} &= \bar{V}_{ext} + \bar{V}_{in} \nonumber \\
    &=\epsilon (\ket{e}\bra{i} + \ket{i} \bra{e}) \nonumber \\
    &+ \Omega (\ket{g}\bra{m} + \ket{m}\bra{g}) - \Delta\omega \ket{m}\bra{m}.
\end{align}
and in the recharging stage we have
\begin{align}\label{vbarr}
    \bar{V}  = \Omega (\ket{g}\bra{m} + \ket{m}\bra{g}) - \Delta\omega \ket{m}\bra{m}.
\end{align}

Once again, considering that $\gamma_{m}^{-} \gg, \gamma_{m}^{+}, \gamma_{e(i)}^{\pm}, \Omega$, we obtain approximate solutions for the time evolution of the population and coherences of level $|m\rangle$. Here, however, we must also assume that $\gamma_{e(i)}^{-} \gg \gamma_m^{+}$. By doing so, we obtain the following master equation for the effective qutrit of levels $\ket{g}$, $\ket{e}$ and $\ket{i}$:
\begin{align}
    \dot{\varrho}(t) &= - i [\bar{V}_{eff}, \varrho(t)] \nonumber \\
    &+ \frac{\Gamma_i^{+}}{2}(2 \ket{i}\bra{g} \varrho(t) \ket{g} \bra{i} - \{\ket{g}\bra{g}, \varrho(t)\}) \nonumber \\
    &+ \frac{\gamma_i^{-}}{2}(2 \ket{g}\bra{i} \varrho(t) \ket{i} \bra{g} - \{\ket{i}\bra{i}, \varrho(t)\}) \nonumber \\
    &+ \frac{\gamma_e^{+}}{2}(2 \ket{e}\bra{g} \varrho(t) \ket{g} \bra{e} - \{\ket{g}\bra{g}, \varrho(t)\}) \nonumber \\
    &+ \frac{\gamma_e^{-}}{2}(2 \ket{g}\bra{e} \varrho(t) \ket{e} \bra{g} - \{\ket{e}\bra{e}, \varrho(t)\}), 
\end{align}
where $\bar{V}_{eff} = p \Delta\omega \ket{g}\bra{g}$, $\Gamma_i^{+} = \gamma_i^{+} + p \gamma_m^{-}$ and $\{A, B\} = AB + BA$.

\subsection{Calculations for the short cycle limit}
In this section, we use the the conditions of the adiabatic elimination of $|m\rangle$ to calculate the thermodynamic quantities of interest for the machine operating in the limit of short cycles ($ \sum_j \gamma_j^{\pm} \tau \ll 1$, where $\tau$ is the duration of the cycle).

For the short cycle, the operational steady state is obtained solving the equation $\rho_{OSS} = \tilde{\rho}_{OSS} + \tau \mathcal{L}(\tilde{\rho}_{OSS})$, where $\tilde{\rho}_{OSS} = U \rho U^{-1}$, $U = -i (\ket{i}\bra{e} + \ket{e}\bra{i}) + \ket{g}\bra{g} + \ket{m}\bra{m}$ and the coupling with the thermal reservoir, $\mathcal{L}(\tilde{\rho}_{OSS})$, is given by equation \eqref{nuappqe}.

In the rotating reference frame, the equation of motion for the populations and for the coherences between levels $\ket{g}$ and $\ket{m}$ are given by
\begin{equation}\label{ggsc}
    i \Omega (r_{mg} - r_{gm}) = \gamma_e^{-} r_i + \gamma_i^{-} r_e - (\gamma_e^{+} + \gamma_i^{+}) r_g,
\end{equation}
\begin{equation}\label{mmsc}
    -i \Omega (r_{mg} - r_{gm}) = \gamma_m^{+} r_e - \gamma_m^{-} r_m,
\end{equation}
\begin{equation}\label{iisc}
    r_i - r_e = \left[\gamma_i^{+}r_g + \gamma_m^{-}r_m - (\gamma_m^{+} + \gamma_i^{-})r_e\right]\tau,
\end{equation}
\begin{equation}\label{eesc}
    r_e - r_i = \left[\gamma_e^{+}r_g - \gamma_e^{-}r_i\right]\tau,
\end{equation}
\begin{equation}\label{gmsc}
    2 i \Omega (r_g - r_m) = (\gamma_e^{+} + \gamma_i^{+} + \gamma_m^{-} + 2 i \Delta\omega) r_{gm},
\end{equation}
where $r_{jk} = \bra{j} \varrho \ket{k}$ and $\Delta \omega = \omega_f - \omega_m$.
Using Eq. \eqref{gmsc} and its complex conjugate, we obtain
\begin{equation}
    r_{mg} - r_{gm} = - \frac{4 i \Omega \alpha}{\alpha^2 + 4\Delta\omega^2}(r_g - r_m).
\end{equation}
Under the adiabatic elimination conditions, this expression becomes
\begin{equation}
    r_{mg} - r_{gm} = - \frac{4 i \Omega}{\gamma_m^{-^2} + 4\Delta\omega^2}\gamma_m^{-}(r_g - r_m).
\end{equation}
Note that, since $r_m \ll r_g$, we can simplify the expression above to
\begin{equation}\label{mggmsc}
    r_{mg} - r_{gm} \approx - \frac{4 i \Omega}{\gamma_m^{-^2} + 4\Delta\omega^2}\gamma_m^{-}r_g.
\end{equation}
Substituting Eq. \eqref{mggmsc} in Eq. \eqref{ggsc}, we obtain
\begin{equation}\label{rg}
    r_g = \frac{\gamma_e^{-} r_i + \gamma_i^{-} r_e}{\Gamma_i^{+} + \gamma_e^{+}},
\end{equation}
where, $\Gamma_i^{+} = \gamma_i^{+} + p \gamma_m^{-}$ and $p = \frac{4 \Omega^2}{\gamma_m^{-^2} + 4\Delta\omega^2} \ll 1$.

Substituting Eq. \eqref{mggmsc} in Eq. \eqref{mmsc}, we obtain
\begin{equation}\label{rm}
    \gamma_m^{-} r_m = \gamma_m^{+} r_e + p \gamma_m^{-} r_g.
\end{equation}
From the Eq. \eqref{eesc}, we obtain
\begin{equation}\label{re}
    r_e = \frac{r_i\left[1 - (\gamma_e^{+} + \gamma_e^{-}) \tau\right] + \gamma_e^{+} \tau}{1 + \gamma_e^{+} \tau}.
\end{equation}

Substituting Eqs. \eqref{rg} and \eqref{re} in Eq. \eqref{iisc} and doing some algebraic manipulation we obtain
\begin{equation}\label{ri}
    r_i = \frac{\Gamma_i^{+} + \gamma_e^{+} - \gamma_e^{+}\gamma_i^{-} \tau}{2(\Gamma_i^{+} + \gamma_e^{+}) +  \gamma_i^{-} + \gamma_e^{-} - \left[\gamma_e^{-}(\Gamma_i^{+} + \gamma_i^{-}) + \gamma_e^{+} \gamma_i^{-}\right]\tau}.
\end{equation}

Using the relations between $r_e$ and $r_i$ and keeping terms up to first order in $\tau$, the ergotropy, given by $\mathcal{E}^{V} = (\omega_i - \omega_e)(r_i - r_e)$, becomes
\begin{equation}\label{ERGsc}
    \mathcal{E}^{SC} = \frac{\omega_i - \omega_e}{2(\Gamma_i^{+} + \gamma_e^{+}) +  \gamma_i^{-} + \gamma_e^{-}} (\Gamma_i^{+} \gamma_e^{-} - \gamma_i^{-}\gamma_e^{+}) \tau.
\end{equation}
Note that, for the ergotropy to be positive it is necessary that $\frac{\Gamma_i^{+}}{\gamma_i{-}} > \frac{\gamma_e^{+}}{\gamma_e^{-}}$. It means that the temperature of the effective reservoir, $T_i$, that appears due to the adiabatic elimination procedure, has to be larger than the temperature of the thermal reservoir, $T$. For the output power, defined as $\mathcal{P}^{SC} = \mathcal{E}^{SC}/\tau$, we have
\begin{equation}
    \mathcal{P}^{SC} = (\omega_i - \omega_e) \frac{\Gamma_i^{+} \gamma_e^{-} - \gamma_i^{-}\gamma_e^{+}}{2(\Gamma_i^{+} + \gamma_e^{+}) +  \gamma_i^{-} + \gamma_e^{-}}.
\end{equation}

The efficiency of the short cycle is defined by
\begin{equation}\label{EFFsc}
    \eta^{SC} = \frac{\mathcal{E}^{SC}}{E_{in}^{SC}},
\end{equation}
where $E_{in}^{SC}$ is the energy injected into the system in the cycle, given by
\begin{align}\label{einscapp}
    E_{in}^{SC} = Max\{&W_{in}^{SC}, W_{in}^{SC} + Q^{SC}, W_{in}^{SC} + Q_{\gamma_m}^{SC}, \nonumber \\
    &W_{in}^{SC}+ Q_{\gamma_i}^{SC}, W_{in}^{SC} + Q_{\gamma_e}^{SC}\}
\end{align}
where $W_{in}^{SC}$ is the work done on the system due to its coupling to the external work source represented by $V_{in}(t)$,
\begin{align}\label{Winsc}
    W_{in}^{SC} &= \operatorname{Tr}\{\tilde{\rho}(\tau) \dot{V}_{in}(\tau)\}  \tau \nonumber \\
    &= \operatorname{Tr}\{\tilde{\rho}(\tau)[i \omega_f \Omega(\ket{g}\bra{m} e^{i \omega_f \tau} - \ket{m}\bra{g} e^{-i \omega_f \tau})]\} \tau \nonumber \\
    &=i \omega_f \Omega (\rho_{mg}e^{i \omega_f \tau} - \rho_{gm} e^{-i \omega_f \tau}) \tau \nonumber \\
    &= i \omega_f \Omega(r_{mg} - r_{gm})\tau.
\end{align}
Substituting Eq. \eqref{mggmsc} in Eq. \eqref{Winsc}, we obtain
\begin{align}\label{Winsc2}
    W_{in}^{SC} &\approx  p \omega_f  \gamma_m^{-} r_g\tau \nonumber \\
    &= p \omega_f \gamma_m^{-} \frac{\gamma_e^{-} + \gamma_i^{-}}{\kappa} \tau,
\end{align}
where $\kappa = 2\left(\Gamma_i^{+} + \gamma_e^{+}\right) + \gamma_i^{-} +  \gamma_e^{-}$. The total heat exchange between the system and the reservoir, $Q_{SC}$ is given by
\begin{align}
    Q_{SC} &= Q_{\gamma_e}^{SC} + Q_{\gamma_i}^{SC} +Q_{\gamma_m}^{SC},
\end{align}
where, $Q_{\gamma_e}^{SC}$ is the heat exchange associated with the transitions between levels $\ket{e}$ and $\ket{g}$
\begin{align}
    Q_{\gamma_e}^{SC} &= \operatorname{Tr}\{\mathcal{L}_{\gamma_e}[\tilde{\rho}][H_0 + V_{in}(\tau)]\} \tau \nonumber \\
    &= \omega_e \tau (\gamma_e^{+} r_g - \gamma_e^{-} r_i) - \gamma_e^{+} \frac{\omega}{2} \tau (r_{mg} +r_{gm}) \nonumber \\
    &=\omega_e \frac{\gamma_e^{+} \gamma_i^{-} - \Gamma_i^{+} \gamma_e^{-}}{\kappa} \tau - p \gamma_e^{+} \Delta \omega \frac{\gamma_i^{-} + \gamma_e^{-}}{\kappa}\tau,
\end{align}
where
\begin{align}\label{rmgrgm}
    r_{mg} + r_{gm} &= \frac{8  \Omega \Delta\omega}{\gamma_m^{-^2} + 4\Delta\omega^2}(r_g - r_m) \nonumber \\
    &\approx \frac{8  \Omega \Delta\omega}{\gamma_m^{-^2} + 4\Delta\omega^2}r_g,
\end{align}
$Q_{\gamma_i}^{SC}$ is the heat exchange associated with the transitions between levels $\ket{i}$ and $\ket{g}$
\begin{align}
    Q_{\gamma_i}^{SC} &= \operatorname{Tr}\{\mathcal{L}_{\gamma_i}[\tilde{\rho}][H_0 + V_{in}(\tau)]\} \tau \nonumber \\
    &= \omega_i \tau (\gamma_i^{+} r_g - \gamma_i^{-} r_e) - \gamma_i^{+} \frac{\omega}{2} \tau (r_{mg} +r_{gm}) \nonumber \\
    &=\omega_i \frac{\gamma_i^{+} \gamma_e^{-} - \gamma_i^{-}( \gamma_e^{+} + p \gamma_m^{-})}{\kappa} \tau \nonumber \\
    &- p \gamma_i^{+} \Delta \omega \frac{\gamma_i^{-} + \gamma_e^{-}}{\kappa}\tau,
\end{align}
and $Q_{\gamma_m}^{SC}$ is the heat exchange associated with the transitions between levels $\ket{m}$ and $\ket{i}$
\begin{align}
    Q_{\gamma_m}^{SC} &=\operatorname{Tr}\{\mathcal{L}_{\gamma_m}[\tilde{\rho}][H_0 + V_{in}(\tau)]\}\tau \nonumber \\
    &= (\omega_i - \omega_m)\tau (\gamma_m^{-} r_m - \gamma_m^{+} r_e) - \gamma_m^{-} \frac{\Omega}{2} \tau (r_{mg} +r_{gm})\nonumber \\
    &=(\omega_i - \omega_m) \tau i \Omega (r_{mg} - r_{gm}) - \gamma_m^{-} \frac{\Omega}{2} \tau (r_{mg} +r_{gm}) \nonumber \\
    &=(\omega_i - \omega_m) p \gamma_m^{-} \frac{\gamma_i^{-} + \gamma_e^{-}}{\kappa}\tau - p \gamma_m^{-} \Delta \omega \frac{\gamma_i^{-} + \gamma_e^{-}}{\kappa}\tau \nonumber \\
    &= (\omega_i - \omega_f) \hskip 0.08cm p \gamma_m^{-} \frac{\gamma_i^{-} + \gamma_e^{-}}{\kappa}\tau.
\end{align}

In the regime of very low temperature, we have that for $\omega_f \geq \omega_i$
\begin{equation}
    E_{in}^{SC} = W_{in}^{SC},
\end{equation}
and the efficiency is given by
\begin{equation}
    \eta_{SC} = \left(\frac{\omega_i - \omega_e}{\omega_f}\right) \frac{\Gamma_i^{+} \gamma_e^{-} - \gamma_i^{-}\gamma_e^{+}}{p \gamma_m^{-}(\gamma_i^{-} + \gamma_e^{-})}.
\end{equation}
We also have that $p \gamma_m^{-} \gg \gamma_{e(i)}^{+}$, $\gamma_j^{-} \approx \gamma_0$, so
\begin{align}
    \eta_{SC} &\approx \left(\frac{\omega_i - \omega_e}{\omega_f}\right) \frac{\gamma_0^e}{\gamma_0^e + \gamma_0^i} \nonumber \\
    &\approx  \left(1 - \frac{\omega_e}{\omega_i}\right)\frac{\gamma_0^e}{\gamma_0^e + \gamma_0^i}\left(\frac{\omega_m}{\omega_i} - \frac{\Delta \omega}{\omega_i}\right).
\end{align}

For $\omega_f < \omega_i$, we have
\begin{equation}
    E_{in}^{SC} = W_{in}^{SC} + Q_m^{SC},
\end{equation}
so the efficiency is given by 
\begin{equation}
    \eta_{SC} = \left(\frac{\omega_i - \omega_e}{\omega_i}\right) \frac{\Gamma_i^{+} \gamma_e^{-} - \gamma_i^{-}\gamma_e^{+}}{p \gamma_m^{-}(\gamma_i^{-} + \gamma_e^{-})}.
\end{equation}
For temperatures low enough,
\begin{equation}
    \eta_{SC} \approx \left(\frac{\omega_i - \omega_e}{\omega_i}\right) \frac{\gamma_0^e}{\gamma_0^e + \gamma_0^i}.
\end{equation}

\end{document}